\documentclass[10pt,letterpaper]{article}
\usepackage[top=0.85in,left=2.75in,footskip=0.75in]{geometry}

\usepackage{amsmath,amssymb}
\usepackage{bm}

\usepackage{changepage}

\usepackage[utf8x]{inputenc}

\usepackage{textcomp,marvosym}

\usepackage{cite}

\usepackage{nameref,hyperref}

\usepackage[right]{lineno}

\usepackage{microtype}
\DisableLigatures[f]{encoding = *, family = * }

\usepackage[table]{xcolor}

\usepackage{array}

\newcolumntype{+}{!{\vrule width 2pt}}

\newlength\savedwidth

\raggedright
\usepackage[aboveskip=1pt,labelfont=bf,labelsep=period,justification=raggedright,singlelinecheck=off]{caption}

\makeatletter
\renewcommand{\@biblabel}[1]{\quad#1.}
\makeatother

\usepackage{lastpage,fancyhdr,graphicx}
\usepackage{epstopdf}
\fancyheadoffset[L]{2.25in}
\fancyfootoffset[L]{2.25in}
\begin{document}
\vspace*{0.2in}

% Title must be 250 characters or less.
\begin{flushleft}
{\Large
\textbf\newline{Nonlinear eco-evolutionary games with global environmental fluctuations and local environmental feedbacks} % Please use "sentence case" for title and headings (capitalize only the first word in a title (or heading), the first word in a subtitle (or subheading), and any proper nouns).
}
\newline
% Insert author names, affiliations and corresponding author email (do not include titles, positions, or degrees).
\\
Yishen Jiang\textsuperscript{1,3},
Xin Wang\textsuperscript{2,3,4,5,6,7*},
Longzhao Liu\textsuperscript{2,3,4,5,6,7},
Ming Wei\textsuperscript{1,3},
Jingwu Zhao\textsuperscript{8},
Zhiming Zheng\textsuperscript{2,3,4,5,6,7,9,10},
Shaoting Tang\textsuperscript{2,3,4,5,6,7,9,10*}
%with the Lorem Ipsum Consortium\textsuperscript{\textpilcrow}
\\
\bigskip
\textbf{1} School of Mathematical Sciences, Beihang University, Beijing 100191, China
\\
\textbf{2} Institute of Artificial Intelligence, Beihang University, Beijing 100191, China
\\
\textbf{3} Key laboratory of Mathematics, Informatics and Behavioral Semantics (LMIB), Beihang University, Beijing 100191, China
\\
\textbf{4} State Key Lab of Software Development Environment (NLSDE), Beihang University, Beijing 100191, China
\\
\textbf{5} Zhongguancun Laboratory, Beijing, P.R.China
\\
\textbf{6} Beijing Advanced Innovation Center for Future Blockchain and Privacy Computing, Beihang University, Beijing 100191, China
\\
\textbf{7} PengCheng Laboratory, Shenzhen 518055, China
\\
\textbf{8} School of Law, Beihang University, Beijing 100191, China
\\
\textbf{9} Institute of Medical Artificial Intelligence, Binzhou Medical University, Yantai 264003, China
\\
\textbf{10} School of Mathematical Sciences, Dalian University of Technology, Dalian 116024, China
\\
\bigskip

% Insert additional author notes using the symbols described below. Insert symbol callouts after author names as necessary.
% 
% Remove or comment out the author notes below if they aren't used.
%
% Primary Equal Contribution Note
%\Yinyang These authors contributed equally to this work.

% Additional Equal Contribution Note
% Also use this double-dagger symbol for special authorship notes, such as senior authorship.
%\ddag These authors also contributed equally to this work.

% Current address notes
%\textcurrency Current Address: Dept/Program/Center, Institution Name, City, State, Country % change symbol to "\textcurrency a" if more than one current address note
% \textcurrency b Insert second current address 
% \textcurrency c Insert third current address

% Deceased author note
%\dag Deceased

% Group/Consortium Author Note
%\textpilcrow Membership list can be found in the Acknowledgments section.

% Use the asterisk to denote corresponding authorship and provide email address in note below.
* wangxin\_1993@buaa.edu.cn(XW); tangshaoting@buaa.edu.cn(ST)

\end{flushleft}
% Please keep the abstract below 300 words
\section*{Abstract}
Environmental changes play a critical role in determining the evolution of social dilemmas in many natural or social systems. Generally, the environmental changes include two prominent aspects: the global time-dependent fluctuations and the local strategy-dependent feedbacks. However, the impacts of these two types of environmental changes have only been studied separately, a complete picture of the environmental effects exerted by the combination of these two aspects remains unclear. Here we develop a theoretical framework that integrates group strategic behaviors with their general dynamic environments, where the global environmental fluctuations are associated with a nonlinear factor in public goods game and the local environmental feedbacks are described by the `eco-evolutionary game'. We show how the coupled dynamics of local game-environment evolution differs in static and dynamic global environments. In particular, we find the emergence of cyclic evolutions of group cooperation and local environment, which forms an interior irregular loop in the phase plane, depending on the relative changing speed of both global and local environments compared to the strategic change. Our results provide important insights toward how diverse evolutionary outcomes could emerge from the nonlinear interactions between strategies and the changing environments.

% Please keep the Author Summary between 150 and 200 words
% Use first person. PLOS ONE authors please skip this step. 
% Author Summary not valid for PLOS ONE submissions.   
\section*{Author summary}
The intricate interplay between strategic behavior and environment is ubiquitous in complex systems of different scales. Previous works mainly focus on one aspect of the environmental changes: either global environment fluctuations that unidirectionally decide the welfare of the evolutionary dynamics, or local environment feedbacks that coevolve with the strategic behavior. Here we develop a theoretical framework that integrates them both in order to obtain a more complete picture of how group cooperation evolves in a general dynamic environment. We show that global environmental fluctuations can fundamentally alter the dynamical predictions of local game-environment evolution. The most interesting finding is the emergence of cyclic evolutions of group cooperation and local environment, which forms an interior irregular loop in the phase plane, depending on the relative changing speed of both global and local environments compared to the strategic change. Our results show how rich dynamical outcomes arise from the interactions between strategic behaviors and their natural or social environments, which has important practical value for solving social dilemmas in an ever-changing world. 

%\linenumbers

% Use "Eq" instead of "Equation" for equation citations.
\section*{Introduction}
%Lorem ipsum dolor sit~\cite{bib1} amet, consectetur adipiscing elit. Curabitur eget porta erat. Morbi consectetur est vel gravida pretium. Suspendisse ut dui eu ante cursus gravida non sed sem. Nullam Eq~(\ref{eqschemeP}) sapien tellus, commodo id velit id, eleifend volutpat quam. Phasellus mauris velit, dapibus finibus elementum vel, pulvinar non tellus. Nunc pellentesque pretium diam, quis maximus dolor faucibus id.~\cite{bib2} Nunc convallis sodales ante, ut ullamcorper est egestas vitae. Nam sit amet enim ultrices, ultrices elit pulvinar, volutpat risus.

Cooperation promotes the emergence of stronger adaptabilities and more abundant functions in many species, forming the very basis of natural systems at different scales\cite{pennisi2009origin, west2003cooperation, crespi2001evolution, riley2002bacteriocins}. However, the `selfish gene’ widely exists in all biosystems where individuals always make rational choices based on their own benefits\cite{hamilton1963evolution, fehr2003nature, traulsen2010human}, giving rise to social dilemmas of non-cooperation\cite{dawes1980social}. Naturally, understanding why the persistent cooperation occurs ubiquitously, on what conditions a stable cooperation could be maintained or promoted, and how the cooperative behavior evolves under natural selections has long been the core objective of evolutionary game theory \cite{nowak2006evolutionary, hauert2006evolutionary, nowak2004emergence, wu2013adaptive}. In particular, many different mechanisms have been proposed to address the well-known Prisoner’s Dilemma and The Tragedy of the Commons \cite{nash1950bargaining, feeny1990tragedy}, for instance, kin selection\cite{hamilton1964genetical}, direct reciprocity and indirect reciprocity\cite{nowak1998evolution, fu2008reputation, ohtsuki2009indirect}, punishment and reward \cite{fehr2002altruistic, chen2015competition, rand2011evolution}, spatial reciprocity\cite{hauert2004spatial, wakano2009spatial, szolnoki2010reward} and group selection\cite{wilson1994reintroducing}. More realistically, the heterogeneities of the players are also taken into account, such as network topology\cite{gomez2011evolutionary}, selective participation mechanism\cite{hauert2006evolutionary, hauert2002volunteering}, wealth-based selection \cite{chen2016individual}, and the recently studied higher-order interactions\cite{alvarez2021evolutionary}.

While early evolutionary game approach typically focuses on the internal properties of replicator dynamics\cite{nowak2006evolutionary}, assuming that the strategic interactions happen in a fixed environment, the impact of a dynamic environment is ignored. Accordingly, coevolutionary games that incorporate the evolution process of an exogenous environment have been largely studied\cite{hauert2006evolutionary, frank2019foundations}. Coevolution rules introduce the environment-related characteristics into the game, for instance, the interaction network, the size of population, the mobility, aging and reputation of players, which also evolve in time and could affect the evolutionary outcome of strategies\cite{perc2010coevolutionary}. Further, the ecological factors in microbial systems are abstracted into a global time-dependent environment\cite{gokhale2016eco}. Such global environmental changes, reflecting the periodic ecological fluctuations or the rapid ecological perturbations, modify the payoffs of the game through a time-varying function and show highly complex impacts on the evolution of group cooperation and on the evolutionary balance of phenotypes\cite{gokhale2016eco, kleshnina2022shifts}.

Note that coevolutionary games only consider the feedback from the global environment, the dynamics of which is independent from the strategic interactions. However, the bi-directional feedbacks between strategies and the environment are identified in wide a range of real-world systems\cite{cao2021eco, raymond2012dynamics, cortez2020destabilizing}. In microbial systems, cooperation often arises due to the secretion or the release of extracellular enzymes, extracellular antibiotic compounds and etc, which influences the local environmental state that will, in turn, alter the incentive for public goods production\cite{riley2002bacteriocins, west2003cooperation}. Likewise, in modern society, decision-making dynamics of competitive cognitions can reshape the public opinion environment, especially with the rise of large-scale social networks, and this shared media atmosphere in turn affects the benefit of social discussion in decision-making process\cite{liu2021co}. In fact, the interactions between online public discourse and external political environment can lead to the emergence of polarized echo chambers\cite{wang2020public}, which has aroused great concern in recent years\cite{jones2022spatial}. Similar coupled dynamics can also be obtained in psychological–economic systems and social-ecological systems\cite{akccay2018collapse, cortez2020destabilizing,  chen2018punishment}, relating to a number of big challenges such as global climate change, overfishing, anti-vaccine problems, pandemic prevention and control\cite{liu2018evolutionary, pauly2002towards, chen2019imperfect, yong2021noncompliance}.

To characterize such complex feedback loops, an emerging theory of `eco-evolutionary games' is proposed recently\cite{tilman2021evolution, liu2022coevolution}. In eco-evolutionary game, the strategic behaviors change the state of the environment, while in turn the environment alters the payoff structure of the game, driving the replicator dynamics with a strategy-dependent feedback-evolving game\cite{wang2020eco}. Abundant evolutionary outcomes are observed under such framework. Beginning from the simplest form, a two-player game coupled with linear environmental feedback can already generate the persistent oscillation of both population cooperations and environmental states\cite{weitz2016oscillating}. Similar persistent cycles can also occur in asymmetric games with heterogeneous environments\cite{wu2014social, hauert2019asymmetric, tilman2020evolutionary}. As a meaningful extension, a multi-player game coupled with asymmetrical environmental feedback identifies the threshold of the feedback speed that can yield oscillatory convergence to persistent cooperation, highlighting the importance of time-scales\cite{shao2019evolutionary}. An innovative manifold control approach is further proposed to steer the eco-evolutionary dynamics to a desired direction\cite{wang2020steering}.

Despite the progress, current models consider only one aspect of the dynamic environments, either global environmental changes which are time-dependent or local environmental feedbacks which are strategy-dependent. In real-world systems, however, these two aspects often coexist and exert complex forces on the strategic evolutions. For example, in the context of the COVID-19 pandemic, cooperation in public health measures has strong impacts on disease spreading, and vice versa\cite{yong2021noncompliance}. Beyond this eco-evolution, the seasonal fluctuations of virus's transmissibility also alter the payoffs of the strategic behaviors\cite{shek2003epidemiology}. In crowdsourcing projects, cooperations can emerge from the asymmetric incentive feedback which results in a local feedback loop\cite{shao2019evolutionary, brabham2013crowdsourcing}. Meanwhile, the periodic fluctuations of the global economic environment\cite{gilchrist2012credit}, which obviously cannot be influenced by the strategic behaviors within the projects, could affect the synergy and discounting of the group payoffs\cite{hauert2006synergy}. Understanding the evolution of cooperation in such complex systems thus has profound practical significance, calling for a framework that could unveil the complete picture of the environmental effects where both global and local environmental dynamics are incorporated.

Here we develop a theoretical framework that integrates group strategic behaviors with their general dynamic environments. In specific, the global environmental fluctuation that influences the group benefit is characterized by a nonlinear factor in public goods game, while the local environmental feedback driven by an asymmetric incentive mechanism is described by the `eco-evolutionary game'. Of particular interest, we show how global environmental fluctuations alter the dynamical predictions of the group cooperation in local feedback-evolving game. Different from the previous observation that local game-environment dynamics could eventually evolve to a stable interior fixed point where cooperation and defection coexist, we find the emergence of cyclic evolutions of group cooperation and local environment under a periodically changing global environment. Surprisingly, such eco-evolutionary dynamics could form an interior closed yet irregular orbit in the phase plane, depending on the relative time-scale of both global and local environments versus strategic changes. Our results provide novel insights toward how complex group behaviors emerge from the nonlinear interactions between strategies and dynamic environments, especially in a non-autonomous system, which is important for understanding the evolution of social dilemmas in a changing world.

%\begin{eqnarray}
%\label{eqschemeP}
%	\mathrm{P_Y} = \underbrace{H(Y_n) - H(Y_n|\mathbf{V}^{Y}_{n})}_{S_Y} + \underbrace{H(Y_n|\mathbf{V}^{Y}_{n})- H(Y_n|\mathbf{V}^{X,Y}_{n})}_{T_{X\rightarrow Y}},
%\end{eqnarray}

\section*{Materials and methods}
%\subsection*{Etiam eget sapien nibh}

% For figure citations, please use "Fig" instead of "Figure".
% Place figure captions after the first paragraph in which they are cited.
\begin{figure}[!ht]
\includegraphics[width=0.95\linewidth]{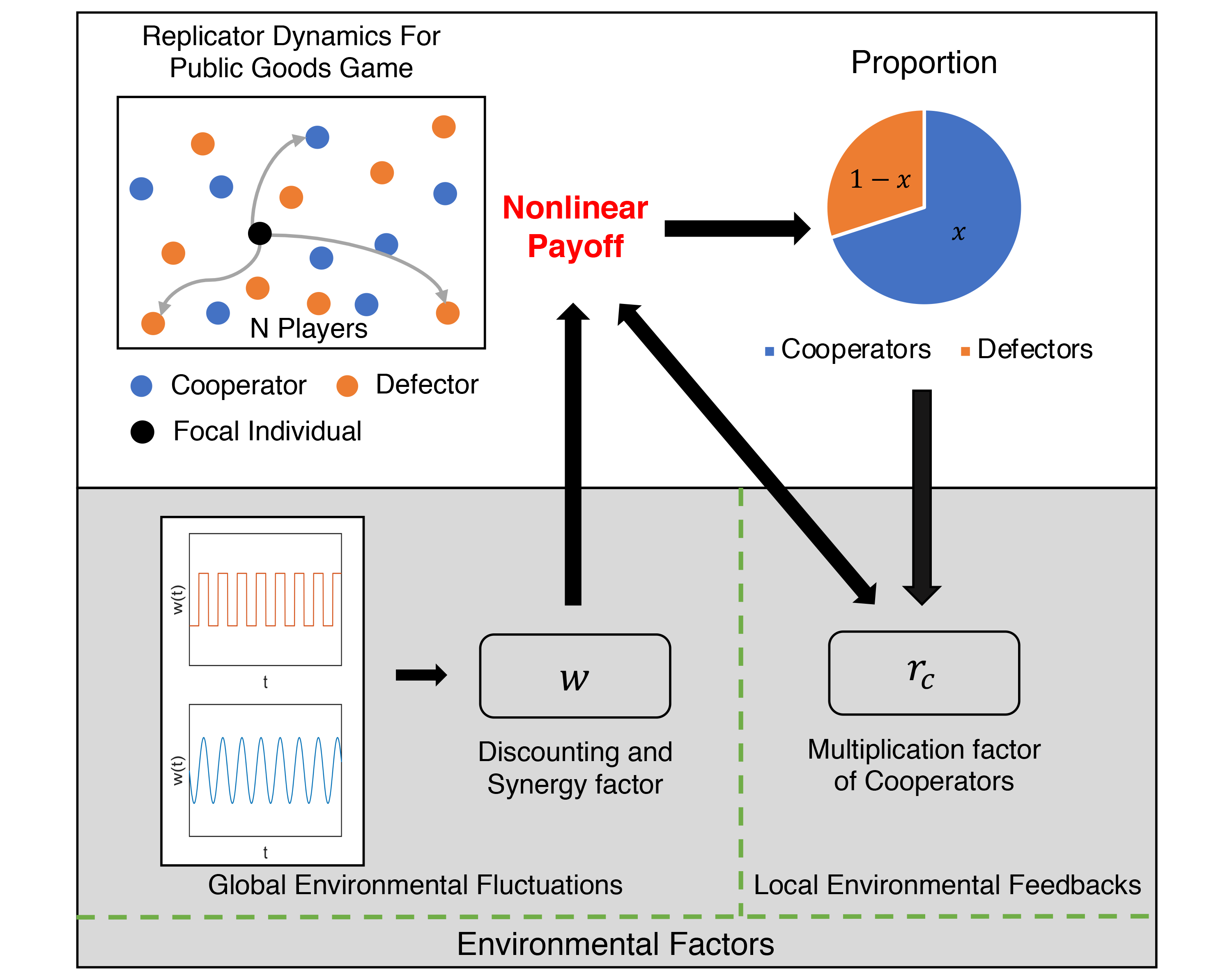}
\caption{{\bf Schematic of the eco-evolutionary games with general dynamic environments.}
(Top) The group strategic behaviors are described by a nonlinear evolutionary public goods game. (Top, Bottom) The influence of environmental changes consist of two prominent aspects: the global environmental fluctuations that directly affect the synergy and discounting of the group payoffs, characterized by the nonlinear factor $w$, and the asymmetric environmental feedbacks that drive the local strategy-dependent feedback-evolving game, characterized by the multiplication factor of cooperators $r_{c}$.}
\label{fig:1}
\end{figure}

In order to study the evolution of group cooperation, which is ubiquitous in microbial systems and in human society, we consider a modified nonlinear public goods game (PGG) among a well-mixed infinitely large population. Each of the $N$ participants can choose to be a cooperator contributing $c$ to the public pool, or a defector reaping without sowing. In classic PGG, after the total contribution is multiplied by the multiplication factor $r$, the total benefit is equally distributed to each participant. However, it has been obtained that group cooperation often emerges due to the the existence of preferential access to the valuable common good or other extra incentives for cooperators \cite{shao2019evolutionary, wang2020steering}, which brings about asymmetric payoff structures for cooperators and defectors and further drives the local feedback-evolving game. Here we distinguish multiplication factors of cooperators and defectors as $r_{c}$ and $r_{d}$, respectively. In addition, the actual benefits provided by cooperators may depend nonlinearly on the number of cooperators and on the total investments, the former is common in biology and the later has been largely revealed in economics \cite{kollock1998social}. Hence, we adopt the modeling idea proposed in \cite{hauert2006synergy} and capture such nonlinearity by a nonlinear factor $w$.

Accordingly, the payoffs for each defector and cooperator in a group with $k$ cooperators, $P_{d}(k)$ and $P_{c}(k)$, are
\begin{eqnarray}
\label{eq1}
{P_d}\left( k \right) = \frac{{{r_d}c}}{N}\left( {1 + w + {w^2} +  \cdots  + {w^{k - 1}}} \right) = \frac{{{r_d}c}}{N}\frac{{1 - {w^k}}}{{1 - w}},
\end{eqnarray}
\begin{eqnarray}
\label{eq2}
{P_c}\left( k \right) = \frac{{{r_c}c}}{N}\left( {1 + w + {w^2} +  \cdots  + {w^{k - 1}}} \right) - c = \frac{{{r_c}c}}{N}\frac{{1 - {w^k}}}{{1 - w}} - c,
\end{eqnarray}
such that the benefits created by each additional cooperator are rescaled, either synergistically enhanced when $w>1$, or discounted when $w<1$. Note that when $w=1$, the classic linear PGG can be recovered. In a population with a fraction $x$ of cooperators, for any focal individual, the probability that $k$ out of $N-1$ other participants are cooperators is
\begin{eqnarray}
\label{eq3}
\left( {\begin{array}{*{20}{c}}
{N - 1}\\
k
\end{array}} \right){x^k}{\left( {1 - x} \right)^{N - 1 - k}},
\end{eqnarray}
The average fitness of defectors and cooperators, $f_{d}$ and $f_{c}$, are thus given by
\begin{eqnarray}
\label{eq4}
&& {f_d} = \sum\limits_{k = 0}^{N - 1} {\left( {\begin{array}{*{20}{c}}
{N - 1}\\
k
\end{array}} \right){x^k}{{\left( {1 - x} \right)}^{N - 1 - k}}{P_d}\left( k \right)} \notag\\
&& = \frac{{{r_d}c}}{{N\left( {1 - w} \right)}}\left[ {1 - {{\left( {1 - x + wx} \right)}^{N - 1}}} \right],
\end{eqnarray}
\begin{eqnarray}
\label{eq5}
&& {f_c} = \sum\limits_{k = 0}^{N - 1} {\left( {\begin{array}{*{20}{c}}
{N - 1}\\
k
\end{array}} \right){x^k}{{\left( {1 - x} \right)}^{N - 1 - k}}{P_c}\left( {k + 1} \right)}\notag\\
&& = c\left\{ {\frac{{{r_c}}}{{N\left( {1 - w} \right)}}\left[ {1 - w{{\left( {1 - x + wx} \right)}^{N - 1}}} \right] - 1} \right\}.
\end{eqnarray}

For simplicity and without loss of generality, we specify that the contribution of each cooperator is $1$. The changes in the fraction of cooperation over time, namely the evolution of group cooperation, is then described by the replicator dynamics:
\begin{eqnarray}
\label{eq6}
&& \dot x = x\left( {{f_c} - \bar f} \right) = x\left( {1 - x} \right)\left( {{f_c} - {f_d}} \right)\notag\\
&& = x{\mkern 1mu} \left( {1 - x} \right){\mkern 1mu} \left( {\frac{{{r_c}{\mkern 1mu} \left( {w{\mkern 1mu} {{\left( {w{\mkern 1mu} x - x + 1} \right)}^{N - 1}} - 1} \right)}}{{N{\mkern 1mu} \left( {w - 1} \right)}} - 1 - \frac{{{r_d}\left( {{{\left( {w{\mkern 1mu} x - x + 1} \right)}^{N - 1}} - 1} \right)}}{{N{\mkern 1mu} \left( {w - 1} \right)}}} \right),
\end{eqnarray}
where $\bar f = x{f_c} + \left( {1 - x} \right){f_d}$ is the average fitness of the population.

Further, we introduce two prominent aspects of the environmental influence into the framework: the global environmental fluctuations and the local environmental feedbacks. Generally, the global environment significantly affects the total and marginal benefits of group cooperation. For instance, the companies tend to increase the salary and recruit more employees in bull markets, adopting aggressive expansion strategies as each additional staff could provide much more returns, while in contrary, they are more likely to cut salaries and reduce stuff in bear markets. Such global environmental fluctuations could naturally be reflected and characterized by the nonlinear factor $w$: a large synergy effect of cooperation corresponds to a good global environment where $w>1$, while the discounting of cooperation is related to a bad global environment where $w<1$. Therefore, we simply use a time-dependent function $w=w\left(t\right)$ to depict changes of the global environment over time.

In addition, the local environmental feedback arised from the asymmetric incentive mechanism, which is strategy-dependent, is characterized by the dynamics of cooperator's multiplication factor $r_c$ following \cite{shao2019evolutionary}:
\begin{eqnarray}
\label{eq7}
{\dot r_c} = \epsilon \left( {{r_c} - \alpha } \right)\left( {\beta  - {r_c}} \right)f\left( {x,{r_c}} \right),
\end{eqnarray}
where $\epsilon>0$ denotes the relative changing speed of $r_c$ compared with $x$. Due to limited resources in local environment, $r_c$ is confined to the range  \([\alpha,\beta]\) and we have $1 < \alpha  < \beta  < N$ according to the social dilemma in PGG. Moreover, $f\left( {x,{r_c}} \right)$ describes the asymmetric feedback mechanism in the model, whose sign determines the increase or decrease in $r_c$. To mimic the fact that in social-economic systems such as crowdsourcing projects, the authoritative organizer could distribute the total payoffs for cooperators and defectors in an effort to promote collaboration, we assume
\begin{eqnarray}
\label{eq8}
f\left( {x,{r_c}} \right) =  - x{f_c} + \theta \left( {1 - x} \right){f_d},
\end{eqnarray}
where $\theta>0$ denotes the distribution ratio of the expected total payoffs of the cooperators and defectors. Such local feedback-evolving dynamics could facilitate cooperation by increasing $r_c$ when $x$ is small, while in turn, a relatively large $x$ leads to the decrease of cooperator's rewards, subjecting to the law of diminishing marginal utility. Besides, to reduce model complexity, the defectors’ multiplication factor $r_{d}$ is set to be constant and ${r_d} \le \alpha$.

The complete modeling framework is illustrated by Fig. \ref{fig:1}. The dynamics of our nonlinear eco-evolutionary game with global environmental fluctuations and local environmental feedbacks, which describes the complex group strategic behaviors in general dynamic environments, can thus be written as

\begin{eqnarray}
\label{eq9}
\left\{ \begin{array}{l}
\dot x = {\mkern 1mu} x{\mkern 1mu} \left( {1 - x} \right){\mkern 1mu} \left( {\frac{{{r_c}{\mkern 1mu} \left( {w{\mkern 1mu} \left( t \right){{\left( {w\left( t \right){\mkern 1mu} x - x + 1} \right)}^{N - 1}} - 1} \right)}}{{N{\mkern 1mu} \left( {w\left( t \right) - 1} \right)}} - 1 - \frac{{{r_d}\left( {{{\left( {w{\mkern 1mu} \left( t \right)x - x + 1} \right)}^{N - 1}} - 1} \right)}}{{N{\mkern 1mu} \left( {w\left( t \right) - 1} \right)}}} \right)\\
{{\dot r}_c} = \epsilon \left( {{r_c} - \alpha } \right){\mkern 1mu} \left( {\beta  - {r_c}} \right){\mkern 1mu} \left( { - x{\mkern 1mu} \left( {\frac{{{r_c}{\mkern 1mu} \left( {w\left( t \right){\mkern 1mu} {{\left( {w\left( t \right){\mkern 1mu} x - x + 1} \right)}^{N - 1}} - 1} \right)}}{{N{\mkern 1mu} \left( {w\left( t \right) - 1} \right)}} - 1} \right) }\right.\\
\left. { + \frac{{{r_d}\theta \left( {1 - x} \right){\mkern 1mu} \left( {{{\left( {w{\mkern 1mu} \left( t \right)x - x + 1} \right)}^{N - 1}} - 1} \right)}}{{N{\mkern 1mu} \left( {w\left( t \right) - 1} \right)}}} \right){\mkern 1mu}
\end{array} \right..
\end{eqnarray}

% Results and Discussion can be combined.
\section*{Results}
%Nulla mi mi, venenatis sed ipsum varius, Table~\ref{table1} volutpat euismod diam. Proin rutrum vel massa non gravida. Quisque tempor sem et dignissim rutrum. Lorem ipsum dolor sit amet, consectetur adipiscing elit. Morbi at justo vitae nulla elementum commodo eu id massa. In vitae diam ac augue semper tincidunt eu ut eros. Fusce fringilla erat porttitor lectus cursus, vel sagittis arcu lobortis. Aliquam in enim semper, aliquam massa id, cursus neque. Praesent faucibus semper libero.

\subsection*{Nonlinear dynamics of local game-environment evolution in static global environments}
\begin{figure}[ht]
\includegraphics[width=0.95\linewidth]{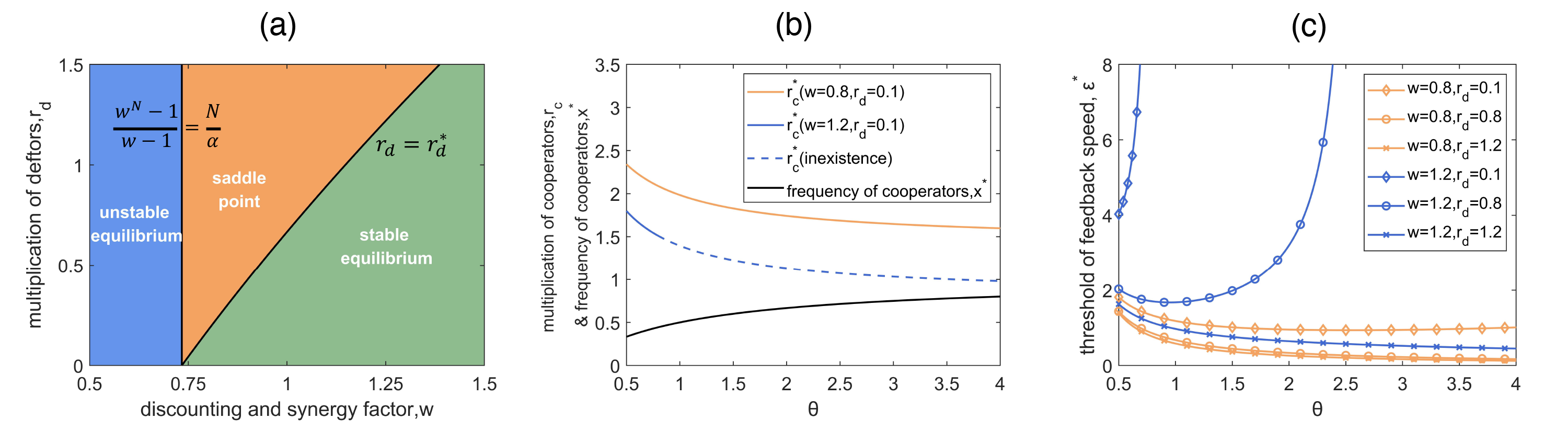}
\caption{{\bf Effects of varying modeling parameters on equilibrium states of local game-environment evolution in static global environments.}
Panel (a) shows the combined influence of $w$ and $r_{d}$ on the stability of the boundary fixed point $\left(1,\alpha\right)$. Panel (b) and (c) show the trends of the interior fixed point $\left(x^{*},r_{c}^{*}\right)$ and the threshold of the relative feedback speed $\epsilon^{*}$ as $\theta$ changes. In all panels, $N=4$, $\alpha=1.5$, $\beta=3.5$.}
\label{fig:2}
\end{figure}

We first study the nonlinear dynamics of local game-environment evolution under static global environments where $w(t)$ is a fixed constant. Correspondingly, the eco-evolutionary game described by Eq. \ref{eq9} degenerates to an autonomous system. There are totally seven possible fixed points in the system, six of which are on the boundary and the remaining one is an interior equilibrium point (See detailed proof of the stability of all seven fixed points in Appendix A). Only two boundary fixed points are possible to be stable: (i) $\left(x^*=0\right)$ is always stable, leading to a full defection among population which also occurs in the classic PGG. (ii) $\left( {{x^*} = 1,r_c^* = \alpha } \right)$ is stable only if ${\mkern 1mu} \frac{{\alpha {\mkern 1mu} \left( {{w^N} - 1} \right)}}{{N{\mkern 1mu} \left( {w - 1} \right)}} > 1$ and ${r_d}$ is smaller than the threshold $r_d^*= \frac{{\alpha \left( {{w^N} - 1} \right) - N\left( {w - 1} \right)}}{{\left( {{w^{N - 1}} - 1} \right)}} $. Under this circumstance, the system evolves to the full cooperation with a minimum value $\alpha$ of cooperators’ multiplication factor. The phase diagram of the stability of this fixed point with respect to $w$ and $r_{d}$ is shown in Fig. \ref{fig:2}(a). Not surprisingly, the full cooperation situation tends to emerge from a healthy global environment with larger $w$ and smaller $r_d$.

Of particular interest, we analyze on what condition the interior equilibrium point in the system
\begin{eqnarray}
\label{eq10}
({x^*} = \frac{\theta }{{\theta  + 1}},r_c^* = \frac{{N\left( {w - 1} \right) + {r_d}\left( {{{\left( {w{x^*} - {x^*} + 1} \right)}^{N - 1}} - 1} \right)}}{{w{{\left( {w{x^*} - {x^*} + 1} \right)}^{N - 1}} - 1}}),
\end{eqnarray}
could be stable, bringing about the coexistence of cooperators and defectors with an intermediate local environmental state. As shown by Eq. \ref{eq10}, the final frequency of cooperators $x^*$ is solely determined by $\theta$, the distribution ratio of cooperator’s and defector’s total payoffs. As $\theta$ increases, $x^*$ also increases whereas the stable multiplication factor of cooperators $r_c^*$ decreases (Fig. \ref{fig:2}(b)). Such changing trends are intuitive in many teamworks: an unskilled project, which has a relatively larger $\theta$, often requires more participants with lower benefits. In contrary, a skilled project such as scientific collaboration, the decrease of $\theta$ results in small research teams with higher reward for each member. Since $\alpha<r_{c}^{*}<\beta$, we first identify the existence condition of this equilibrium point
\begin{eqnarray}
\label{eq11}
\max \left\{ {0,\frac{{\alpha \left( {w{{\left( {\frac{{w\theta  + 1}}{{\theta  + 1}}} \right)}^{N - 1}} - 1} \right) - N\left( {w - 1} \right)}}{{{{\left( {\frac{{w\theta  + 1}}{{\theta  + 1}}} \right)}^{N - 1}} - 1}}} \right\} \le {r_d}\notag\\
\le \min \left\{ {\alpha ,\frac{{\beta \left( {w{{\left( {\frac{{w\theta  + 1}}{{\theta  + 1}}} \right)}^{N - 1}} - 1} \right) - N\left( {w - 1} \right)}}{{{{\left( {\frac{{w\theta  + 1}}{{\theta  + 1}}} \right)}^{N - 1}} - 1}}} \right\}.
\end{eqnarray}

\begin{figure}[ht]
\includegraphics[width=0.95\linewidth]{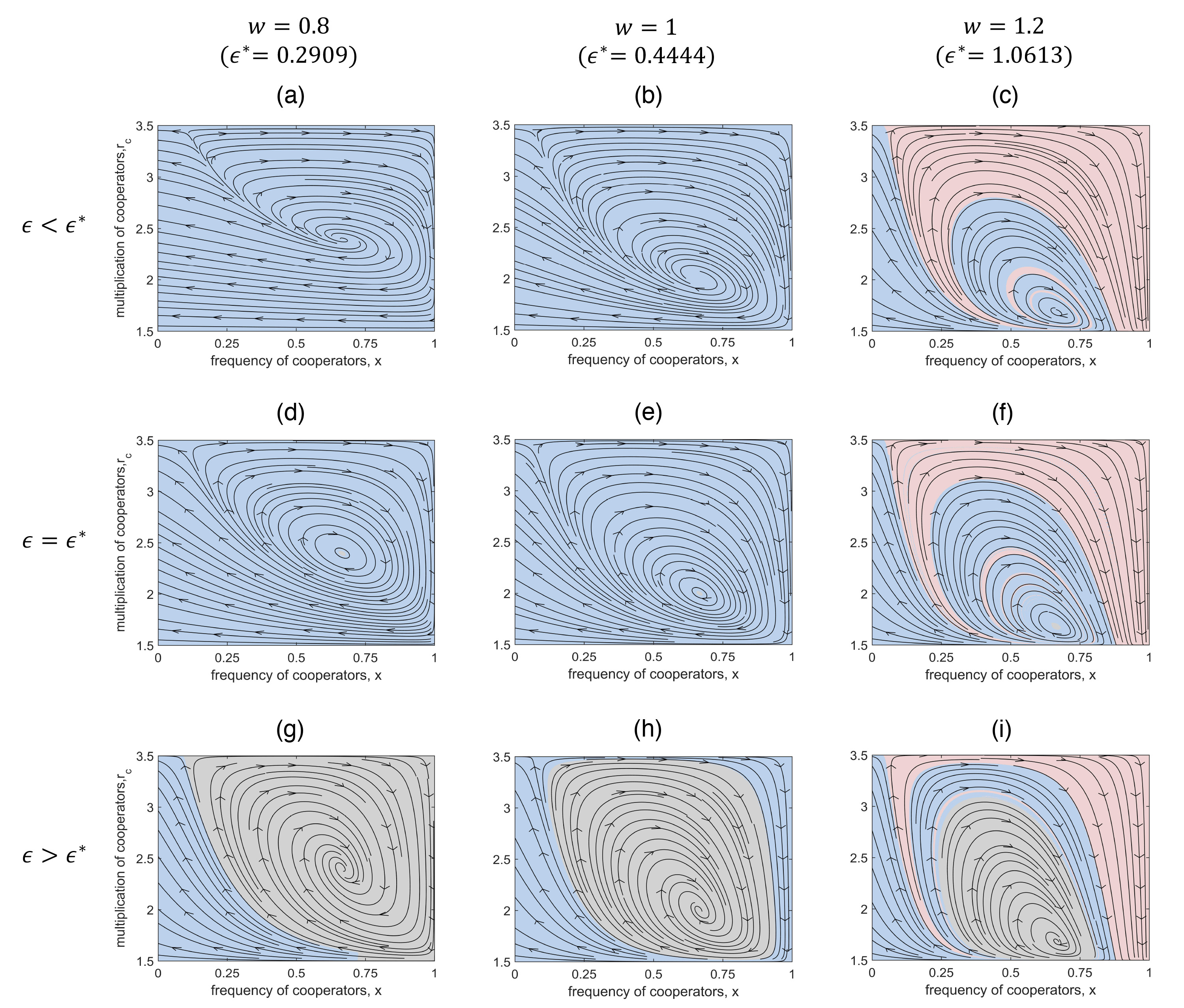}
\caption{{\bf Phase plane dynamics of local game-environment ($\bm{x-r_{c}}$) evolution with $\bm{N=4}$, $\bm{\alpha=1.5}$, $\bm{\beta=3.5}$, $\bm{\theta=2}$, $\bm{r_{d}=1}$.}
We set $w=0.8$, $1$, $1.2$ to describe the scenarios of discounting, linear, and synergy PGG in each column, and the corresponding $\epsilon^*$ are $0.2909$, $0.4444$, $1.0613$, respectively. We therefore choose $\epsilon=0.0909, 0.2444, 0.8613$ for the first row and $\epsilon=0.7909, 0.9444, 1.5613$ for the third row. The blue, pink and gray areas represent the basin of attraction of different fixed points $\left(x^{*}=0\right), \left(x^{*}=1, r_{c}^{*}=\alpha\right)$ and (${x^*} = \frac{\theta }{{\theta  + 1}},r_c^* = \frac{{N\left( {w - 1} \right) + {r_d}\left( {{{\left( {w{x^*} - {x^*} + 1} \right)}^{N - 1}} - 1} \right)}}{{w{{\left( {w{x^*} - {x^*} + 1} \right)}^{N - 1}} - 1}}$), respectively.}
\label{fig:3}
\end{figure}

Further, the existing interior fixed point is stable only if the relative feedback speed of the local environment $\epsilon$ exceeds a threshold $\epsilon^{*}$, which can be written as follows:
\begin{eqnarray}
\label{eq12}
{\epsilon ^*} = \frac{{\left( {r_c^*w - {r_d}} \right)\left( {N - 1} \right){{\left( {w\theta  + 1} \right)}^{N - 2}}\left( {w - 1} \right)}}{{\left( {r_c^* - \alpha } \right)\left( {\beta  - r_c^*} \right)\left[ {w{{\left( {w\theta  + 1} \right)}^{N - 1}} - {{\left( {1 + \theta } \right)}^{N - 1}}} \right]}}.
\end{eqnarray}
Fig. \ref{fig:2}(c) shows how $\epsilon^{*}$ varies as $w$, $r_d$ and $\theta$ change. In a less optimistic global environment where the group benefits are discounted ($w=0.8$), $\epsilon^{*}$ gently decreases as $\theta$ increases. In a booming global environment where the group benefits are synergistically enhanced ($w=1.2$), however, the trends become complicated, depending on the multiplication factor of defectors $r_d$. When $r_d$ becomes smaller, $\epsilon^{*}$ sharply increases as the interior fixed point tends to disappear, the latter is shown by the dash line in Fig. \ref{fig:2}(b).

In Fig. \ref{fig:3}, we show phase dynamics of the local game-environment evolution under different static global environments. We set $w=0.8$, $1$, $1.2$ to mimic the scenarios of discounting, linear, and synergy PGG in each column, respectively. Throughout the paper, we fix $N=4$, $\alpha=1.5$, $\beta=3.5$. Other parameters are $\theta=2$ and $r_{d}=1$. Accordingly, we can calculate the corresponding threshold $\epsilon^{*}$ using Eq. \ref{eq12} and select parameters $\epsilon$ that are smaller than, equal to or larger than $\epsilon^{*}$, respectively. Consistent with our theoretical predictions, the persistent coexistence of cooperators and defectors only occurs when $\epsilon>\epsilon^{*}$ in all three scenarios, where the system oscillatorily converges to the interior equilibrium state. In particular, the group cooperation may only arise from the local asymmetrical environmental feedbacks that are quick and timely enough, especially when the global environment is relatively poor (Fig. \ref{fig:3}(a)(d)(g) and Fig. \ref{fig:3}(b)(e)(h)). Moreover, we find that a good global environment significantly promotes the emergence of full cooperation. An important insight is that a slower local environmental feedback, though obstructs the emergence of interior equilibrium, could indeed increase the basin of attraction of full cooperation (Fig. \ref{fig:3}(c)(f)(i)). In addition, as the benefits brought by the global environment $w$ increase, the stable coexistence of cooperation and defection arise with less asymmetric incentive feedback $r_c^*$, leading to a reduction in the basin of attraction of the interior equilibrium. Our results show non-negligible role and complicated joint influence of global environmental state and local environmental feedback on group strategy evolutions.

\subsection*{Nonlinear dynamics of local game-environment evolution in dynamic global environments}
Further, we study how dynamic global environments affect the local game-environment evolution. Specifically, we focus on periodic global environmental fluctuations, which is typical and widespread in various complex systems. For instance, the daily cycle of sunlight, the seasonal fluctuations of ecological characteristics, the cyclical economic crisis and etc. Here we consider two types of periodic changes: discrete shifts modeled by a piecewise function $w_{1}\left(t\right)$ and continuous fluctuations modeled by a continuous function $w_{2}\left(t\right )$, which are simply given by
\begin{eqnarray}
\label{eq13}
{w_{\rm{1}}}\left( t \right) = \left\{ {\begin{array}{*{20}{c}}
{1.2}&{\left[ {\dfrac{t}{T}} \right] = 2n}\\
{0.8}&{\left[ {\dfrac{t}{T}} \right] = 2n+1}
\end{array}} \right.,n = 0,1,2, \cdots  \cdots.
\end{eqnarray}\\
\begin{eqnarray}
\label{eq14}
{w_2}\left( t \right) ={1 - 0.5\sin \left( {at + \delta } \right)}.
\end{eqnarray}
where $a$ decides the time scales of global environmental fluctuations and $\delta$ modulates the initial phase. Note that $w_2(t)$ is bounded in $[0.5, 1.5]$ in order to balance the degree of discounting and synergy. In addition, for the convenience of comparison, we let the periods of $w_1\left(t\right)$ and $w_2\left(t\right)$ be equal, i.e., $T = \frac{{\rm{\pi }}}{a}$.

\begin{figure}[ht]
\includegraphics[width=0.95\linewidth]{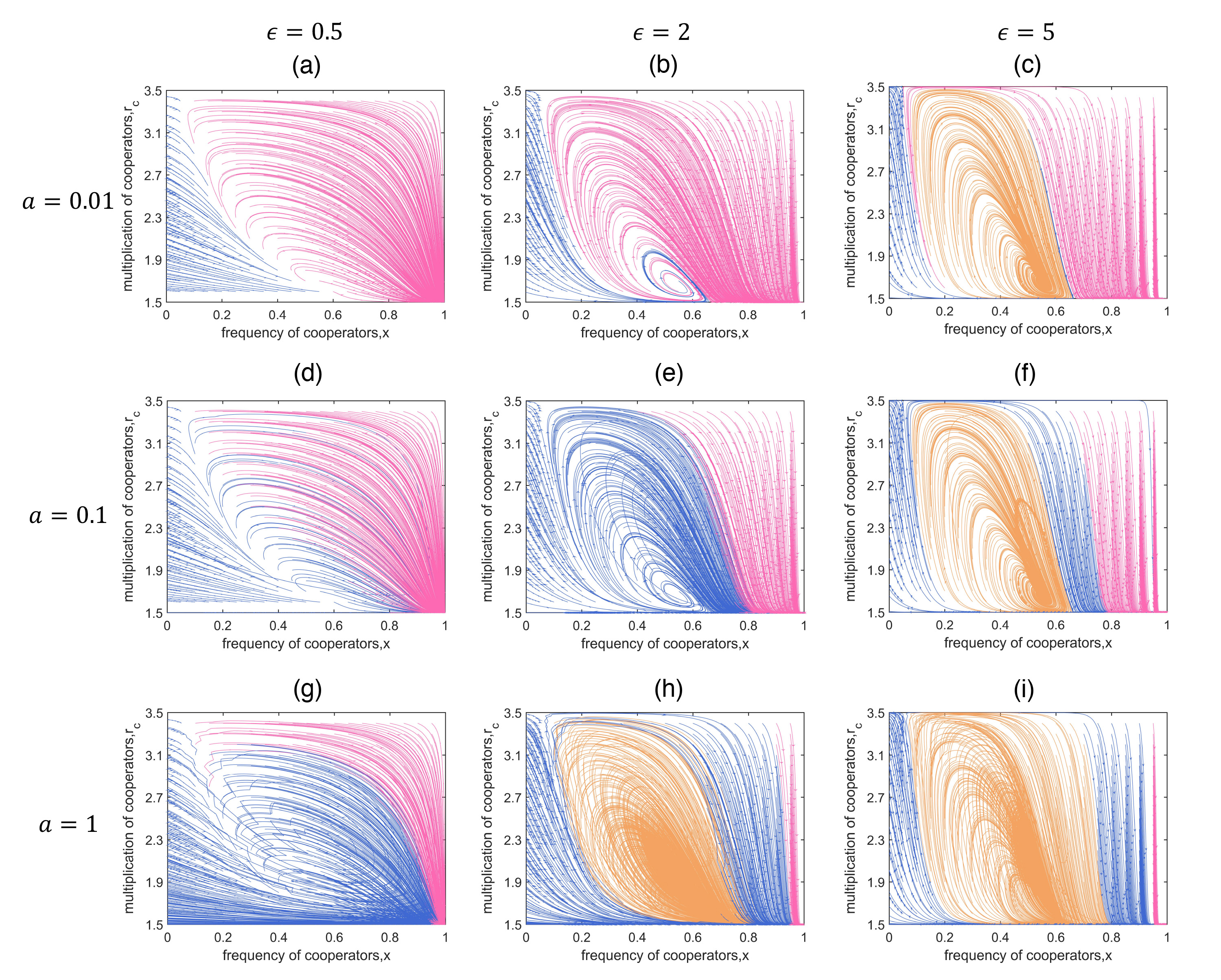}
\caption{{\bf Local game-environment evolution in a discretely varying global environment.}
We use a periodic piecewise function, $w_{1}\left(t\right)$, to describe the global environmental fluctuations. We uniformly select initial points on the $x-r_{c}$ plane and plot the corresponding dynamical trajectories by numerically solving Eq. \ref{eq9}. Trajectories that eventually evolve to $\left(x^{*}=0\right)$, $\left(x^{*}=1, r_{c}^{*}=\alpha\right)$ or circulate along an interior closed orbit are distinguished by blue, pink and orange, respectively. In all panels, $N=4$, $\alpha=1.5$, $\beta=3.5$, $\theta=1.2$, $r_{d}=0.7$.}
\label{fig:4}
\end{figure}

In Fig. \ref{fig:4}, we show the dynamical trajectories of local game-environment evolutions under a discretely varying global environments $w=w_{1}\left(t\right)$. The initial points are uniformly selected on the phase plane and the trajectories are calculated numerically by Eq. \ref{eq9}. We fix $\theta=1.2$ and $r_{d}=0.7$. The thresholds of the local environmental feedback speed corresponding to the two values in $w_{1}\left(t\right)$ thus can be obtained using Eq. \ref{eq12}, which are $\epsilon_{1}^*=0.5944$ and $\epsilon_{2}^*=2.3578$. Therefore, we choose $\epsilon=0.5, 2, 5$, which satisfies $0.5 < \epsilon_{1}^* < 2 <\epsilon_{2}^* <5$, such that our analysis could contain all the possible situations. Furthermore, considering the fact that the global environmental changes are commonly much slower than the strategy evolution, we set the relative changing speed $a=0.01$, $0.1$, $1$ and the corresponding periods are $2T=200\pi$, $20\pi$ and $2\pi$, respectively. We find that the global environmental fluctuations could fundamentally alter the dynamical predictions of the group cooperation in local feedback-evolving game. When $\epsilon=0.5$, the local eco-evolutionary dynamics evolves either to full defection where $x=0$ (blue trajectories) or to full cooperation where $x=1$ and $r_c=\alpha$ (pink trajectories). When $\epsilon=2$, however, a new evolutionary outcome emerges when the global environment changes fastly: the local game-environment dynamics will eventually circulate along an interior closed orbit (orange trajectories). More specifically, different from the phenomenon that the system oscillatorily converges to the interior equilibrium as we observed in static global environments, here we find that a periodically changing global environment could lead to the emergence of cyclic evolutions of group cooperation and local environment. When $\epsilon=5$, such cyclic evolution can even emerge from various time-scales of the global environmental changes.

\begin{figure}[!ht]
\includegraphics[width=0.95\linewidth]{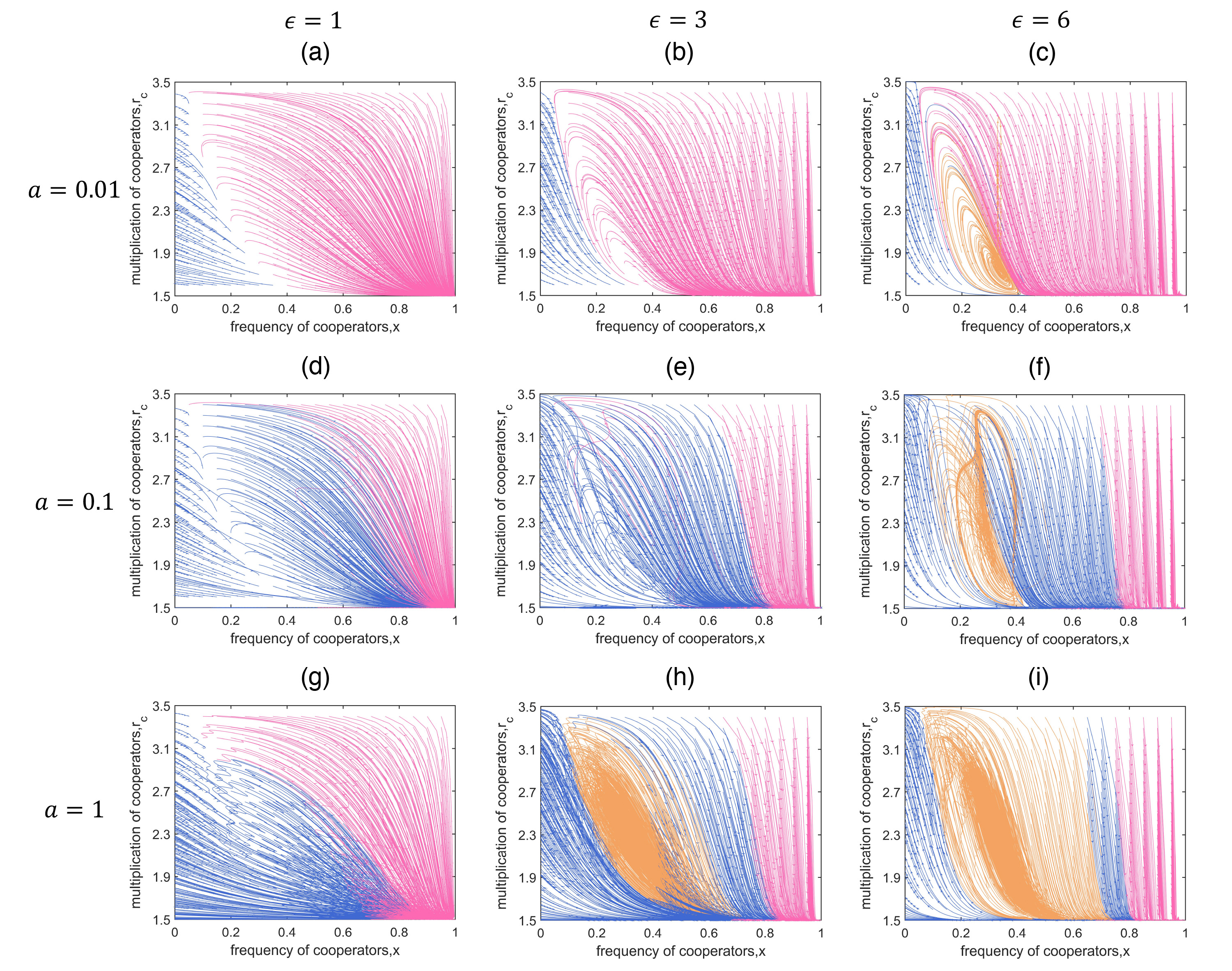}
\caption{{\bf Local game-environment evolution in a continuously changing global environment.}
We use a periodic continuous function, $w_{2}\left(t\right)$, to describe the global environmental fluctuations. Similarly, trajectories that eventually evolve to $\left(x^{*}=0\right)$, $\left(x^{*}=1, r_{c}^{*}=\alpha\right)$ or circulate along an interior closed orbit are distinguished by blue, pink and orange, respectively. In all panels, $N=4$, $\alpha=1.5$, $\beta=3.5$, $\theta=0.5$, $r_{d}=0.6$, $\delta = - \frac{{\rm{\pi }} }{2}$.}
\label{fig:5}
\end{figure}

Similar evolutionary trends can also be observed in a continuously changing global environment $w=w_{2}\left(t\right)$ (Fig. \ref{fig:5}). We set $\theta=0.5$, $ r_{d}=0.6$, $\delta =- \frac{{\rm{\pi }} }{2}$. The thresholds corresponding to the maximum and minimum values of $w_{2}\left(t\right)$ are $\epsilon_{max}=5.2864$ and $\epsilon_{min}=1.4528$. Thus we choose $\epsilon=1, 3, 6$ such that $1 < \epsilon_{min} < 3 < \epsilon_{max} < 6$. Likewise, the population will be occupied either by defectors or cooperators when $\epsilon$ is small. When $\epsilon$ is at an intermediate value, the cyclic evolution along an interior closed orbit arise from a rapidly changing global environment. When $\epsilon$ is large, such interior closed orbit occurs at all global time-scales. In addition, we confirm that the emergence of cyclic evolution is robust on different multiplication factors of defector $r_d$ and on initial phase $\delta$ (see Appendix B). It is noteworthy that these two factors have strong and complicated impacts on the range of initial values that could eventually stabilize to full cooperation ($x^{*}=1$ and $r_{c}^{*}=\alpha$) or to cyclic evolutions.

\begin{figure}[!ht]
\includegraphics[width=0.95\linewidth]{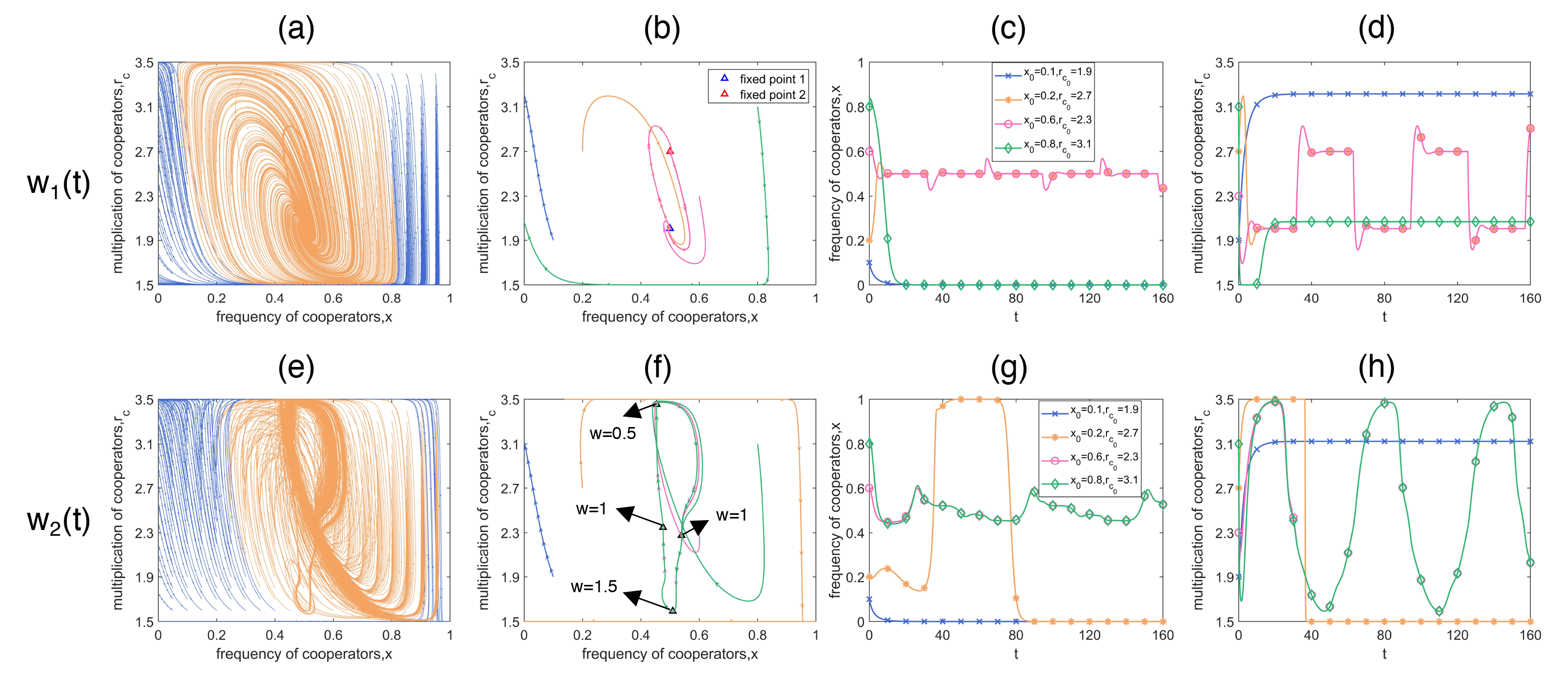}
\caption{{\bf Emergence of cyclic evolutions of group cooperation and local environment under periodically changing global environments given by $\bm{w_{1}\left(t\right)}$ and $\bm{w_{2}\left(t\right)}$.}
The first column shows local game-environment evolutions similar to Fig. \ref{fig:4} and Fig. \ref{fig:5}. The second column displays four typical dynamic trajectories in detail, particularly the interior closed yet irregular orbits. The last two columns present time evolutions of the frequency of cooperators $x$ and the multiplication factor of cooperators $r_c$ under different initial conditions, corresponding to the colored trajectories in the second column. Parameters are $N=4$, $\alpha=1.5$, $\beta=3.5$, $\theta=1$, $r_{d}=1.2$, $a=0.1$, $\epsilon=4$, $\delta=0$.}
\label{fig:6}
\end{figure}

These results just show how diverse evolutionary group behaviors emerge from the complex interactions between strategies and general dynamic environments, especially highlight the important role of relative time-scales of both global and local environments versus strategic changes ($a$ and $\epsilon$). Lastly, we concentrate on the newly discovered phenomenon that the local game-environment dynamics can evolve cyclically under different periodic global environments (Fig. \ref{fig:6}). Due to the complexity of non-autonomous system and the lack of theoretical insight, we display in detail the phase plane dynamics of local game-environment evolution, the typical dynamic trajectories with different initial conditions, and the corresponding time evolutions of the frequency of cooperators $x$ and the multiplication factor of cooperators $r_c$, in order to provide a clear view of the cyclic evolutions. We set $\theta=1$, $r_{d}=1.2$, $a=0.1$, $\delta = 0$ and choose $\epsilon=4$ which is larger than the maximum threshold of $\epsilon^*$ with regard to $w_1(t)$ and $w_2(t)$. Importantly, we find that the reason for the formation of cyclic evolution under discrete global environment $w_1$ is that the two fixed points are in the mutual attraction domains (Fig. \ref{fig:6}(b)). In the shown cases, the initial points first evolve to the stable fixed point $\left( {{x^*} = 0.5, {r_c^*}=2.0047} \right)$ where $w=1.2$ within time $T$. Subsequently, the global environment changes and fixed point $1$ becomes a new initial point, which is captured by another stable fixed point $\left( {{x^*} = 0.5, {r_c^*}=2.6996} \right)$ corresponding to $w=0.8$. Similarly, in the next $T$ time, the evolution is directed to fixed point $1$ again, resulting in the formation of an interior closed orbit. The periodic fluctuations of group cooperation and the local environment are shown in Fig. \ref{fig:6}(c-d). In a continuously changing global environment described by $w_2(t)$, however, the fixed point also varies over time which is unreachable for the evolutionary process. Surprisingly, the interior closed yet irregular orbit still emerges. Such periodic evolution is actually in line with the intuitions from real world in a way that the group cooperation will neither disappear completely nor always be maintained at the highest level in many complex systems, for instance, the seasonal oscillating dynamics of the COVID-19 infection \cite{weitz2020awareness, merow2020seasonality, bergman2020oscillations}, and the dynamic adjustment of big companies in the economic cycles. We interpret our results applying limit thought: the evolution is always moving towards the nearest stable fixed point in each small time interval, and the dynamical trajectories are approximately the combinations of all such short paths connecting head to tail, the limit of which finally converge to an closed yet irregular orbit as we observed. A simple verification for the described thought using a discretely changing global environment, which is characterized by a segmented function of four segments, is provided in Appendix B.

\section*{Conclusion and discussion}

%CO\textsubscript{2} Maecenas convallis mauris sit amet sem ultrices gravida. Etiam eget sapien nibh. Sed ac ipsum eget enim egestas ullamcorper nec euismod ligula. Curabitur fringilla pulvinar lectus consectetur pellentesque. Quisque augue sem, tincidunt sit amet feugiat eget, ullamcorper sed velit. 

%Sed non aliquet felis. Lorem ipsum dolor sit amet, consectetur adipiscing elit. Mauris commodo justo ac dui pretium imperdiet. Sed suscipit iaculis mi at feugiat. Ut neque ipsum, luctus id lacus ut, laoreet scelerisque urna. Phasellus venenatis, tortor nec vestibulum mattis, massa tortor interdum felis, nec pellentesque metus tortor nec nisl. Ut ornare mauris tellus, vel dapibus arcu suscipit sed. Nam condimentum sem eget mollis euismod. Nullam dui urna, gravida venenatis dui et, tincidunt sodales ex. Nunc est dui, sodales sed mauris nec, auctor sagittis leo. Aliquam tincidunt, ex in facilisis elementum, libero lectus luctus est, non vulputate nisl augue at dolor. For more information, see \nameref{S1_Appendix}.
How rich dynamical outcomes arise from the interactions between strategic behaviors and their natural or social environments is one of the fundamental questions in many complex systems across disciplines. On one hand, the global environment unidirectionally changes the total welfare of the evolutionary dynamics. One the other hand, the local environment provides frequency-dependent feedback that modifies the payoff structure of the game dynamics, while in turn the strategies taken by individuals can also reshape the state of the local environment over time. Particularly, two important theoretical frameworks that describe the latter feedback loop are proposed: the stochastic game\cite{hilbe2018evolution, su2019evolutionary}, and the eco-evolutionary game\cite{weitz2016oscillating, tilman2020evolutionary, shao2019evolutionary, wang2020steering}. Stochastic games introduce game transition mechanisms to depict the discrete changes of the external environments, i.e., the cooperation behavior in the current game can affect the game that individuals play in the next period. Furthermore, eco-evolutionary game theory characterizes the continuous environmental changes coupled with strategic interactions via a set of ordinary differential equations. However, previous models exclusively focus on one aspect of the environmental changes. To obtain a more complete picture of the group behavioral evolutions in a general dynamic environment, we develop a modeling framework that integrates the more complicated influence exerted by both global environmental fluctuations and local environmental feedbacks.    

Real interactions between strategic behaviors and their environments are commonly nonlinear. Our analysis shows how this nonlinearity, relating to the state of the global environment and the marginal benefits provided by the cooperators, affect the local game-environment evolution. We find that in a static global environment, regardless of the scenarios of discounting, linear, and synergy, the persistent coexistence of cooperation and defection only emerges if the relative feedback speed of the local environment exceeds a certain threshold, breaking the `Tragedy of the Commons'. The nonlinear factor, however, could determine the occurrence of full cooperation and influence the attraction basin of the stable interior equilibrium.  

As the influence of dynamic global environments on local eco-evolutionary game is the primary focus of our model, our results emphatically show how the periodic global environmental fluctuations fundamentally alter the evolutionary outcomes of the group cooperation. The most intriguing finding is the emergence of an interior closed yet irregular orbit in local game-environment phase plane, leading to cyclic evolutions of group cooperation and local environment, which is firstly discovered in multi-player situation and qualitatively in line with the oscillating dynamics in two-player games \cite{weitz2016oscillating, tilman2020evolutionary}. Further, we unveil that this new dynamical phenomenon can be intuitively understood as a limit of continuously converging process of the dynamical paths confined by the fixed points. Last and most important, our theoretical framework highlight the crucial role of relative time-scales of both global and local environments compared to strategic interactions.

To sum up, our model provides profound insight into how diverse group behaviors, especially oscillating convergence and cyclic evolutions, can emerge from the nonlinear interactions between strategies and dynamic environments, which has important practical value for solving social dilemmas in an ever-changing world. For instance, in social-economic system, the companies engaging staff can be mapped into the local environment, the state of which coevolves with individual strategies\cite{bowles2003co}. Outside, the trends of the corresponding industry, which can largely determine the marginal benefits of strategic behaviors, can be modeled as the global environment. Our framework also has great potential in studying more complicated cases as well as bigger issues in human society, such as the coevolution of legal environment, new technologies and the economics\cite{aaronson2015trade}. 

Though in this work we only consider the periodic fluctuations of the global environment which are widespread in natural systems, various changing rules can be studied by simply giving different time-evolving functions $w(t)$. Future works may further consider the structured interactions with increasing complexity, and the control strategies in the framework of game-environment dynamics.

\section*{Acknowledgments}
This work is supported by Program of National Natural Science Foundation of China Grant No. 12201026,11871004, 11922102, 62141605 and National Key Research and Development Program of China Grant No. 2018AAA0101100, 2021YFB2700304.

\nolinenumbers

% Either type in your references using
% \begin{thebibliography}{}
% \bibitem{}
% Text
% \end{thebibliography}
%
% or
%
% Compile your BiBTeX database using our plos2015.bst
% style file and paste the contents of your .bbl file
% here. See http://journals.plos.org/plosone/s/latex for 
% step-by-step instructions.
% 
%\begin{thebibliography}{10}

%\bibitem{bib1}
%Conant GC, Wolfe KH.
%\newblock {{T}urning a hobby into a job: how duplicated genes find new
%  functions}.
%\newblock Nat Rev Genet. 2008 Dec;9(12):938--950.

%\bibitem{bib2}
%Ohno S.
%\newblock Evolution by gene duplication.
%\newblock London: George Alien \& Unwin Ltd. Berlin, Heidelberg and New York:
%  Springer-Verlag.; 1970.

%\bibitem{bib3}
%Magwire MM, Bayer F, Webster CL, Cao C, Jiggins FM.
%\newblock {{S}uccessive increases in the %resistance of {D}rosophila to viral
%  infection through a transposon insertion followed by a {D}uplication}.
%\newblock PLoS Genet. 2011 Oct;7(10):e1002337.

%\end{thebibliography}

%\bibliography{ref.bib}

\begin{thebibliography}{10}

\bibitem{pennisi2009origin}
Pennisi E. On the origin of cooperation; 2009.

\bibitem{west2003cooperation}
West SA, Buckling A.
\newblock Cooperation, virulence and siderophore production in bacterial
  parasites.
\newblock Proceedings of the Royal Society of London Series B: Biological
  Sciences. 2003;270(1510):37--44.

\bibitem{crespi2001evolution}
Crespi BJ.
\newblock The evolution of social behavior in microorganisms.
\newblock Trends in ecology \& evolution. 2001;16(4):178--183.

\bibitem{riley2002bacteriocins}
Riley MA, Wertz JE.
\newblock Bacteriocins: evolution, ecology, and application.
\newblock Annual Reviews in Microbiology. 2002;56(1):117--137.

\bibitem{hamilton1963evolution}
Hamilton WD.
\newblock The evolution of altruistic behavior.
\newblock The American Naturalist. 1963;97(896):354--356.

\bibitem{fehr2003nature}
Fehr E, Fischbacher U.
\newblock The nature of human altruism.
\newblock Nature. 2003;425(6960):785--791.

\bibitem{traulsen2010human}
Traulsen A, Semmann D, Sommerfeld RD, Krambeck HJ, Milinski M.
\newblock Human strategy updating in evolutionary games.
\newblock Proceedings of the National Academy of Sciences.
  2010;107(7):2962--2966.

\bibitem{dawes1980social}
Dawes RM.
\newblock Social dilemmas.
\newblock Annual review of psychology. 1980;31(1):169--193.

\bibitem{nowak2006evolutionary}
Nowak MA.
\newblock Evolutionary dynamics: exploring the equations of life.
\newblock Harvard university press; 2006.

\bibitem{hauert2006evolutionary}
Hauert C, Holmes M, Doebeli M.
\newblock Evolutionary games and population dynamics: maintenance of
  cooperation in public goods games.
\newblock Proceedings of the Royal Society B: Biological Sciences.
  2006;273(1600):2565--2571.

\bibitem{nowak2004emergence}
Nowak MA, Sasaki A, Taylor C, Fudenberg D.
\newblock Emergence of cooperation and evolutionary stability in finite
  populations.
\newblock Nature. 2004;428(6983):646--650.

\bibitem{wu2013adaptive}
Wu T, Fu F, Zhang Y, Wang L.
\newblock Adaptive role switching promotes fairness in networked ultimatum
  game.
\newblock Scientific reports. 2013;3(1):1--8.

\bibitem{nash1950bargaining}
Nash~Jr JF.
\newblock The bargaining problem.
\newblock Econometrica: Journal of the econometric society. 1950; p. 155--162.

\bibitem{feeny1990tragedy}
Feeny D, Berkes F, McCay BJ, Acheson JM.
\newblock The tragedy of the commons: twenty-two years later.
\newblock Human ecology. 1990;18(1):1--19.

\bibitem{hamilton1964genetical}
Hamilton WD.
\newblock The genetical evolution of social behaviour. II.
\newblock Journal of theoretical biology. 1964;7(1):17--52.

\bibitem{nowak1998evolution}
Nowak MA, Sigmund K.
\newblock Evolution of indirect reciprocity by image scoring.
\newblock Nature. 1998;393(6685):573--577.

\bibitem{fu2008reputation}
Fu F, Hauert C, Nowak MA, Wang L.
\newblock Reputation-based partner choice promotes cooperation in social
  networks.
\newblock Physical Review E. 2008;78(2):026117.

\bibitem{ohtsuki2009indirect}
Ohtsuki H, Iwasa Y, Nowak MA.
\newblock Indirect reciprocity provides only a narrow margin of efficiency for
  costly punishment.
\newblock Nature. 2009;457(7225):79--82.

\bibitem{fehr2002altruistic}
Fehr E, G{\"a}chter S.
\newblock Altruistic punishment in humans.
\newblock Nature. 2002;415(6868):137--140.

\bibitem{chen2015competition}
Chen X, Szolnoki A, Perc M.
\newblock Competition and cooperation among different punishing strategies in
  the spatial public goods game.
\newblock Physical Review E. 2015;92(1):012819.

\bibitem{rand2011evolution}
Rand DG, Nowak MA.
\newblock The evolution of antisocial punishment in optional public goods
  games.
\newblock Nature communications. 2011;2(1):1--7.

\bibitem{hauert2004spatial}
Hauert C, Doebeli M.
\newblock Spatial structure often inhibits the evolution of cooperation in the
  snowdrift game.
\newblock Nature. 2004;428(6983):643--646.

\bibitem{wakano2009spatial}
Wakano JY, Nowak MA, Hauert C.
\newblock Spatial dynamics of ecological public goods.
\newblock Proceedings of the National Academy of Sciences.
  2009;106(19):7910--7914.

\bibitem{szolnoki2010reward}
Szolnoki A, Perc M.
\newblock Reward and cooperation in the spatial public goods game.
\newblock EPL (Europhysics Letters). 2010;92(3):38003.

\bibitem{wilson1994reintroducing}
Wilson DS, Sober E.
\newblock Reintroducing group selection to the human behavioral sciences.
\newblock Behavioral and brain sciences. 1994;17(4):585--608.

\bibitem{gomez2011evolutionary}
Gomez-Gardenes J, Romance M, Criado R, Vilone D, S{\'a}nchez A.
\newblock Evolutionary games defined at the network mesoscale: The public goods
  game.
\newblock Chaos: An Interdisciplinary Journal of Nonlinear Science.
  2011;21(1):016113.

\bibitem{hauert2002volunteering}
Hauert C, De~Monte S, Hofbauer J, Sigmund K.
\newblock Volunteering as red queen mechanism for cooperation in public goods
  games.
\newblock Science. 2002;296(5570):1129--1132.

\bibitem{chen2016individual}
Chen X, Szolnoki A.
\newblock Individual wealth-based selection supports cooperation in spatial
  public goods games.
\newblock Scientific reports. 2016;6(1):1--8.

\bibitem{alvarez2021evolutionary}
Alvarez-Rodriguez U, Battiston F, de~Arruda GF, Moreno Y, Perc M, Latora V.
\newblock Evolutionary dynamics of higher-order interactions in social
  networks.
\newblock Nature Human Behaviour. 2021;5(5):586--595.

\bibitem{frank2019foundations}
Frank SA.
\newblock Foundations of social evolution.
\newblock In: Foundations of Social Evolution. Princeton University Press;
  2019.

\bibitem{perc2010coevolutionary}
Perc M, Szolnoki A.
\newblock Coevolutionary games—a mini review.
\newblock BioSystems. 2010;99(2):109--125.

\bibitem{gokhale2016eco}
Gokhale CS, Hauert C.
\newblock Eco-evolutionary dynamics of social dilemmas.
\newblock Theoretical Population Biology. 2016;111:28--42.

\bibitem{kleshnina2022shifts}
Kleshnina M, McKerral JC, Gonz{\'a}lez-Tokman C, Filar JA, Mitchell JG.
\newblock Shifts in evolutionary balance of phenotypes under environmental
  changes.
\newblock Royal Society Open Science. 2022;9(11):220744.

\bibitem{cao2021eco}
Cao L, Wu B.
\newblock Eco-evolutionary dynamics with payoff-dependent environmental
  feedback.
\newblock Chaos, Solitons \& Fractals. 2021;150:111088.

\bibitem{raymond2012dynamics}
Raymond B, West SA, Griffin AS, Bonsall MB.
\newblock The dynamics of cooperative bacterial virulence in the field.
\newblock Science. 2012;337(6090):85--88.

\bibitem{cortez2020destabilizing}
Cortez MH, Patel S, Schreiber SJ.
\newblock Destabilizing evolutionary and eco-evolutionary feedbacks drive
  empirical eco-evolutionary cycles.
\newblock Proceedings of the Royal Society B. 2020;287(1919):20192298.

\bibitem{liu2021co}
Liu H, Wang X, Liu L, Li Z.
\newblock Co-evolutionary game dynamics of competitive cognitions and public
  opinion environment.
\newblock Frontiers in Physics. 2021;9:658130.

\bibitem{wang2020public}
Wang X, Sirianni AD, Tang S, Zheng Z, Fu F.
\newblock Public discourse and social network echo chambers driven by
  socio-cognitive biases.
\newblock Physical Review X. 2020;10(4):041042.

\bibitem{jones2022spatial}
Jones MI, Pauls SD, Fu F.
\newblock Spatial Games of Fake News.
\newblock arXiv preprint arXiv:220604118. 2022;.

\bibitem{akccay2018collapse}
Ak{\c{c}}ay E.
\newblock Collapse and rescue of cooperation in evolving dynamic networks.
\newblock Nature communications. 2018;9(1):1--9.

\bibitem{chen2018punishment}
Chen X, Szolnoki A.
\newblock Punishment and inspection for governing the commons in a
  feedback-evolving game.
\newblock PLoS computational biology. 2018;14(7):e1006347.

\bibitem{liu2018evolutionary}
Liu L, Wang S, Chen X, Perc M.
\newblock Evolutionary dynamics in the public goods games with switching
  between punishment and exclusion.
\newblock Chaos: An Interdisciplinary Journal of Nonlinear Science.
  2018;28(10):103105.

\bibitem{pauly2002towards}
Pauly D, Christensen V, Gu{\'e}nette S, Pitcher TJ, Sumaila UR, Walters CJ,
  et~al.
\newblock Towards sustainability in world fisheries.
\newblock Nature. 2002;418(6898):689--695.

\bibitem{chen2019imperfect}
Chen X, Fu F.
\newblock Imperfect vaccine and hysteresis.
\newblock Proceedings of the royal society B. 2019;286(1894):20182406.

\bibitem{yong2021noncompliance}
Yong JC, Choy BK.
\newblock Noncompliance with safety guidelines as a free-riding strategy: an
  evolutionary game-theoretic approach to cooperation during the COVID-19
  pandemic.
\newblock Frontiers in Psychology. 2021;12:729.

\bibitem{tilman2021evolution}
Tilman AR, Vasconcelos VV, Akcay E, Plotkin JB.
\newblock The evolution of forecasting for decision making in dynamic
  environments.
\newblock arXiv preprint arXiv:210800047. 2021;.

\bibitem{liu2022coevolution}
Liu Y, Wu B.
\newblock Coevolution of vaccination behavior and perceived vaccination risk
  can lead to a stag-hunt-like game.
\newblock Physical Review E. 2022;106(3):034308.

\bibitem{wang2020eco}
Wang X, Fu F.
\newblock Eco-evolutionary dynamics with environmental feedback: Cooperation in
  a changing world.
\newblock Europhysics Letters. 2020;132(1):10001.

\bibitem{weitz2016oscillating}
Weitz JS, Eksin C, Paarporn K, Brown SP, Ratcliff WC.
\newblock An oscillating tragedy of the commons in replicator dynamics with
  game-environment feedback.
\newblock Proceedings of the National Academy of Sciences.
  2016;113(47):E7518--E7525.

\bibitem{wu2014social}
Wu T, Fu F, Dou P, Wang L.
\newblock Social influence promotes cooperation in the public goods game.
\newblock Physica A: Statistical Mechanics and its Applications.
  2014;413:86--93.

\bibitem{hauert2019asymmetric}
Hauert C, Saade C, McAvoy A.
\newblock Asymmetric evolutionary games with environmental feedback.
\newblock Journal of theoretical biology. 2019;462:347--360.

\bibitem{tilman2020evolutionary}
Tilman AR, Plotkin JB, Ak{\c{c}}ay E.
\newblock Evolutionary games with environmental feedbacks.
\newblock Nature communications. 2020;11(1):1--11.

\bibitem{shao2019evolutionary}
Shao Y, Wang X, Fu F.
\newblock Evolutionary dynamics of group cooperation with asymmetrical
  environmental feedback.
\newblock EPL (Europhysics Letters). 2019;126(4):40005.

\bibitem{wang2020steering}
Wang X, Zheng Z, Fu F.
\newblock Steering eco-evolutionary game dynamics with manifold control.
\newblock Proceedings of the Royal Society A. 2020;476(2233):20190643.

\bibitem{shek2003epidemiology}
Shek LPC, Lee BW.
\newblock Epidemiology and seasonality of respiratory tract virus infections in
  the tropics.
\newblock Paediatric respiratory reviews. 2003;4(2):105--111.

\bibitem{brabham2013crowdsourcing}
Brabham DC.
\newblock Crowdsourcing.
\newblock Mit Press; 2013.

\bibitem{gilchrist2012credit}
Gilchrist S, Zakraj{\v{s}}ek E.
\newblock Credit spreads and business cycle fluctuations.
\newblock American economic review. 2012;102(4):1692--1720.

\bibitem{hauert2006synergy}
Hauert C, Michor F, Nowak MA, Doebeli M.
\newblock Synergy and discounting of cooperation in social dilemmas.
\newblock Journal of theoretical biology. 2006;239(2):195--202.

\bibitem{kollock1998social}
Kollock P.
\newblock Social dilemmas: The anatomy of cooperation.
\newblock Annual review of sociology. 1998; p. 183--214.

\bibitem{weitz2020awareness}
Weitz JS, Park SW, Eksin C, Dushoff J.
\newblock Awareness-driven behavior changes can shift the shape of epidemics
  away from peaks and toward plateaus, shoulders, and oscillations.
\newblock Proceedings of the National Academy of Sciences.
  2020;117(51):32764--32771.

\bibitem{merow2020seasonality}
Merow C, Urban MC.
\newblock Seasonality and uncertainty in global COVID-19 growth rates.
\newblock Proceedings of the National Academy of Sciences.
  2020;117(44):27456--27464.

\bibitem{bergman2020oscillations}
Bergman A, Sella Y, Agre P, Casadevall A.
\newblock Oscillations in US COVID-19 incidence and mortality data reflect
  diagnostic and reporting factors.
\newblock Msystems. 2020;5(4):e00544--20.

\bibitem{hilbe2018evolution}
Hilbe C, {\v{S}}imsa {\v{S}}, Chatterjee K, Nowak MA.
\newblock Evolution of cooperation in stochastic games.
\newblock Nature. 2018;559(7713):246--249.

\bibitem{su2019evolutionary}
Su Q, McAvoy A, Wang L, Nowak MA.
\newblock Evolutionary dynamics with game transitions.
\newblock Proceedings of the National Academy of Sciences.
  2019;116(51):25398--25404.

\bibitem{bowles2003co}
Bowles S, Choi JK, Hopfensitz A.
\newblock The co-evolution of individual behaviors and social institutions.
\newblock Journal of theoretical biology. 2003;223(2):135--147.

\bibitem{aaronson2015trade}
Aaronson S.
\newblock Why trade agreements are not setting information free: The lost
  history and reinvigorated debate over cross-border data flows, human rights,
  and national security.
\newblock World Trade Review. 2015;14(4):671--700.

\end{thebibliography}

\newpage
\appendix
\section*{Appendix A. Stability of fixed points in  static global environments}
For any given $w$ ($w \ne 1$), the dynamic equations of the eco-evolutionary system can be written as
\begin{eqnarray}
\label{eqA1}
\left\{ \begin{array}{l}
\dot x = {\mkern 1mu} x{\mkern 1mu} \left( {1 - x} \right){\mkern 1mu} \left( {\frac{{{r_c}{\mkern 1mu} \left( {w{\mkern 1mu} {{\left( {w{\mkern 1mu} x - x + 1} \right)}^{N - 1}} - 1} \right)}}{{N{\mkern 1mu} \left( {w - 1} \right)}} - 1 - \frac{{{r_d}\left( {{{\left( {w{\mkern 1mu} x - x + 1} \right)}^{N - 1}} - 1} \right)}}{{N{\mkern 1mu} \left( {w - 1} \right)}}} \right)\\
{{\dot r}_c} = \epsilon \left( {{r_c} - \alpha } \right){\mkern 1mu} \left( {\beta  - {r_c}} \right){\mkern 1mu} \left( { - x{\mkern 1mu} \left( {\frac{{{r_c}{\mkern 1mu} \left( {w{\mkern 1mu} {{\left( {w{\mkern 1mu} x - x + 1} \right)}^{N - 1}} - 1} \right)}}{{N{\mkern 1mu} \left( {w - 1} \right)}} - 1} \right) }\right.\\
\left.{ + \frac{{{r_d}\theta \left( {1 - x} \right){\mkern 1mu} \left( {{{\left( {w{\mkern 1mu} x - x + 1} \right)}^{N - 1}} - 1} \right)}}{{N{\mkern 1mu} \left( {w - 1} \right)}}} \right){\mkern 1mu}
\end{array} \right..
\end{eqnarray}

Jacobian matrix of this system is

\begin{equation}J = \left[ {\begin{array}{*{20}{c}}
\begin{array}{l}
A\\
C
\end{array}&\begin{array}{l}
B\\
D
\end{array}
\end{array}} \right],
\end{equation} where

\begin{eqnarray}
&& A =  \left( {2x - 1} \right){\mkern 1mu} \left( {\frac{{{r_d}\left( {{{\left( {w{\mkern 1mu} x - x + 1} \right)}^{N - 1}} - 1} \right)}}{{N{\mkern 1mu} \left( {w - 1} \right)}} - \frac{{{r_c}{\mkern 1mu} \left( {w{\mkern 1mu} {{\left( {w{\mkern 1mu} x - x + 1} \right)}^{N - 1}} - 1} \right)}}{{N{\mkern 1mu} \left( {w - 1} \right)}} + 1} \right)\notag\\
&& + \frac{{{\mkern 1mu} x\left( {1 - x} \right){\mkern 1mu} \left( {{r_c}w - {r_d}} \right){\mkern 1mu} \left( {N - 1} \right){\mkern 1mu} {\mkern 1mu} {{\left( {w{\mkern 1mu} x - x + 1} \right)}^{N - 2}}}}{N}\notag\\\
&& B = \frac{{x\left( {1 - x} \right)\left( {w{\mkern 1mu} {{\left( {w{\mkern 1mu} x - x + 1} \right)}^{N - 1}} - 1} \right){\mkern 1mu} }}{{N{\mkern 1mu} \left( {w - 1} \right)}}\notag\\\
&& C = \epsilon {\mkern 1mu} \left( {{r_c} - \alpha } \right){\mkern 1mu} \left( {\beta  - {r_c}} \right){\mkern 1mu} \left({ \frac{{{r_c}\left( {1 - w{\mkern 1mu} {{\left( {w{\mkern 1mu} x - x + 1} \right)}^{N - 1}}} \right)}}{{N{\mkern 1mu} \left( {w - 1} \right)}}}\right. \notag\\
&& \left.{+ \frac{{{r_d}\theta {\mkern 1mu} \left( {1 - {{\left( {w{\mkern 1mu} x - x + 1} \right)}^{N - 1}}} \right)}}{{N{\mkern 1mu} \left( {w - 1} \right)}}+ \frac{{{r_d}\theta {\mkern 1mu} \left( {N - 1} \right){\mkern 1mu} \left( {1 - x} \right){\mkern 1mu} {{\left( {w{\mkern 1mu} x - x + 1} \right)}^{N - 2}}}}{N}}\right.\notag\\\
&&\left.{ - \frac{{{r_c}w{\mkern 1mu} x{\mkern 1mu} \left( {N - 1} \right){\mkern 1mu} {{\left( {w{\mkern 1mu} x - x + 1} \right)}^{N - 2}}}}{N} + 1 }\right)\notag\\\
&& D = \epsilon \left( {2{r_c} - \alpha  - \beta } \right){\mkern 1mu} \left( {x{\mkern 1mu} \left( {\frac{{{r_c}{\mkern 1mu} \left( {w{\mkern 1mu} {{\left( {w{\mkern 1mu} x - x + 1} \right)}^{N - 1}} - 1} \right)}}{{N{\mkern 1mu} \left( {w - 1} \right)}} - 1} \right)}\right.\notag\\\
&& \left.{ + \frac{{{r_d}\theta {\mkern 1mu} \left( {x - 1} \right){\mkern 1mu} \left( {{{\left( {w{\mkern 1mu} x - x + 1} \right)}^{N - 1}} - 1} \right)}}{{N{\mkern 1mu} \left( {w - 1} \right)}}} \right){\mkern 1mu}\notag\\\
&& +\frac{{\epsilon {\mkern 1mu} x{\mkern 1mu} \left( {1 - w{\mkern 1mu} {{\left( {w{\mkern 1mu} x - x + 1} \right)}^{N - 1}}} \right){\mkern 1mu} \left( {{r_c} - \alpha } \right){\mkern 1mu} \left( {\beta  - {r_c}} \right)}}{{N{\mkern 1mu} \left( {w - 1} \right)}}
\end{eqnarray}

Solving $\dot x=0$ and ${\dot r_c=0}$, we can derive the following fixed points.

(1) ${x^*} = 0$

  \begin{equation}
\begin{aligned}
J\left( {{x^*} = 0,{r_c}} \right) = \left[ {\begin{array}{*{20}{c}}
{\frac{{{r_c}}}{{{\mkern 1mu} N}} - 1}&0\\
{\epsilon {\mkern 1mu} \left( {{r_c} - \frac{3}{2}} \right){\mkern 1mu} \left( {\frac{{\rm{7}}}{{\rm{2}}} - {r_c}} \right){\mkern 1mu} \left( {\frac{{{\mkern 1mu} {r_d}\theta \left( {N - 1} \right)}}{N} - \frac{{{r_c}}}{N} + 1} \right)}&0
\end{array}} \right]
\end{aligned}
  \end{equation}

Eigenvalues are ${\lambda _{\rm{1}}}{\rm{ = }}\frac{{{r_c} - \left( {N + 1} \right)}}{N} < 0$ (since ${r_c} \le \alpha  < N$), ${\lambda _2} = 0$. So this fixed point is stable.

(2) ${x^ * } = 1,r_c^ *  = \alpha $

  \begin{equation}
\begin{aligned}
J\left( {1,\alpha } \right) = \left( {\begin{array}{*{20}{c}}
{\frac{{{r_d}\left( {{w^{N - 1}} - 1} \right) - \alpha \left( {{w^N} - 1} \right)}}{{N{\mkern 1mu} \left( {w - 1} \right)}} + 1}&0\\
0&{\epsilon {\mkern 1mu} \left( {\beta  - \alpha } \right){\mkern 1mu} \left( {1 - \frac{{\alpha {\mkern 1mu} \left( {{w^N} - 1} \right)}}{{N{\mkern 1mu} \left( {w - 1} \right)}}} \right)}
\end{array}} \right)
\end{aligned}
  \end{equation}

Eigenvalues are ${\lambda _1} = \frac{{{r_d}\left( {{w^{N - 1}} - 1} \right) - \alpha \left( {{w^N} - 1} \right)}}{{N{\mkern 1mu} \left( {w - 1} \right)}} + 1$ and ${\lambda _2} = \epsilon {\mkern 1mu} \left( {\beta  - \alpha } \right){\mkern 1mu} \left( {1 - \frac{{\alpha {\mkern 1mu} \left( {{w^N} - 1} \right)}}{{N{\mkern 1mu} \left( {w - 1} \right)}}} \right)$. When ${\mkern 1mu} \frac{{\alpha {\mkern 1mu} \left( {{w^N} - 1} \right)}}{{N{\mkern 1mu} \left( {w - 1} \right)}} > 1$ and ${r_d} < \frac{{\alpha \sum\limits_{k = 0}^{N - 1} {{w^k}}  - N}}{{\sum\limits_{k = 0}^{N - 2} {{w^k}} }} = \frac{{\alpha \left( {{w^N} - 1} \right) - N\left( {w - 1} \right)}}{{\left( {{w^{N - 1}} - 1} \right)}} = r_d^*$, we have ${\lambda _1} < 0$ and ${\lambda _2} < 0$ and this fixed point is stable.

(3) ${x^ * } = 1,r_c^ *  = \beta $

  \begin{equation}
\begin{aligned}
J\left( {1,\beta } \right){\rm{ = }}\left( {\begin{array}{*{20}{c}}
{\frac{{{r_d}\left( {{w^{N - 1}} - 1} \right)}}{{N{\mkern 1mu} \left( {w - 1} \right)}} - \frac{{\beta {\mkern 1mu} \left( {{w^N} - 1} \right)}}{{N{\mkern 1mu} \left( {w - 1} \right)}} + 1}&0\\
0&{\epsilon {\mkern 1mu} \left( {\beta  - \alpha } \right){\mkern 1mu} \left( {\frac{{\beta {\mkern 1mu} \left( {{\mkern 1mu} {w^N} - 1} \right)}}{{N{\mkern 1mu} \left( {w - 1} \right)}} - 1} \right)}
\end{array}} \right)
\end{aligned}
  \end{equation}

Eigenvalues are ${\lambda _1} = \frac{{{r_d}\left( {{w^{N - 1}} - 1} \right)}}{{N{\mkern 1mu} \left( {w - 1} \right)}} - \frac{{\beta {\mkern 1mu} \left( {{w^N} - 1} \right)}}{{N{\mkern 1mu} \left( {w - 1} \right)}} + 1$ and ${\lambda _2} = \epsilon {\mkern 1mu} \left( {\beta  - \alpha } \right){\mkern 1mu} \left( {\frac{{\beta {\mkern 1mu} \left( {{\mkern 1mu} {w^N} - 1} \right)}}{{N{\mkern 1mu} \left( {w - 1} \right)}} - 1} \right)$.
We can conclude that it is impossible for ${\lambda _1} $ and ${\lambda _2}$ to be negative simultaneously, the proof of which is as follows: If ${\lambda _2} < 0$, we have $\frac{{\beta {\mkern 1mu} \left( {{\mkern 1mu} {w^N} - 1} \right)}}{{N{\mkern 1mu} \left( {w - 1} \right)}} < 1$. Furthermore, if ${\lambda _1} < 0$, we have $\frac{{{r_d}\left( {{w^{N - 1}} - 1} \right)}}{{N{\mkern 1mu} \left( {w - 1} \right)}} < \frac{{\beta {\mkern 1mu} \left( {{w^N} - 1} \right)}}{{N{\mkern 1mu} \left( {w - 1} \right)}} -1 < 0$. Then we have ${r_d} < 0$, which is contradicted with the real situation ${r_d} \geq 0$. Thus, there is at least one eigenvalue no less than $0$, which indicates that the fixed point is unstable.

(4) ${x^ * } = 1,r_c^* = \frac{{N\left( {w - 1} \right)}}{{{w^N} - 1}}$ (If $\alpha  < r_c^* < \beta $, the fixed point is in domain of definition)

\begin{equation}
\begin{aligned}
J\left( {{x^*},r_c^*} \right) = \left[ {\begin{array}{*{20}{c}}
A&0\\
C&D
\end{array}} \right]
\end{aligned}
  \end{equation}
where
  \begin{equation}
\begin{aligned}
A &= \frac{{{r_d}{\mkern 1mu} \left( {{w^{N - 1}} - 1} \right)}}{{N{\mkern 1mu} \left( {w - 1} \right)}} - \frac{{r_c^*{\mkern 1mu} \left( {{w^N} - 1} \right)}}{{N{\mkern 1mu} \left( {w - 1} \right)}} + 1 = \frac{{{r_d}{\mkern 1mu} \left( {{w^{N - 1}} - 1} \right)}}{{N{\mkern 1mu} \left( {w - 1} \right)}} > 0\\
C &= \epsilon {\mkern 1mu} \left( {\alpha  - {r_c}} \right){\mkern 1mu} \left( {\beta  - {r_c}} \right){\mkern 1mu} \left( {\frac{{r_c^*{\mkern 1mu} \left( {{w^N} - 1} \right)}}{{N{\mkern 1mu} \left( {w - 1} \right)}} + \frac{{{r_d}\theta {\mkern 1mu} \left( {{w^{N - 1}} - 1} \right)}}{{N{\mkern 1mu} \left( {w - 1} \right)}} + \frac{{r_c^*{w^{N - 1}}{\mkern 1mu} \left( {N - 1} \right)}}{N} - 1} \right)\\
D &= \frac{{\epsilon \left( {\alpha  - r_c^*} \right){\mkern 1mu} \left( {\beta  - r_c^*} \right){\mkern 1mu} \left( {{w^N} - 1} \right)}}{{N{\mkern 1mu} \left( {w - 1} \right)}} < 0
\end{aligned}
  \end{equation}

Eigenvalues are ${\lambda _1} = \frac{{{r_d}{\mkern 1mu} \left( {{w^{N - 1}} - 1} \right)}}{{N{\mkern 1mu} \left( {w - 1} \right)}} > 0$ and ${\lambda _2} = \frac{{\epsilon \left( {\alpha  - r_c^*} \right){\mkern 1mu} \left( {\beta  - r_c^*} \right){\mkern 1mu} \left( {{w^N} - 1} \right)}}{{N{\mkern 1mu} \left( {w - 1} \right)}} < 0$. This fixed point is a saddle point.

(5) ${x^*} = \frac{{{{\left( {\frac{{\alpha  - {r_d} + N{\mkern 1mu} \left( {w - 1} \right)}}{{\alpha w - {r_d}}}} \right)}^{\frac{{\rm{1}}}{{N - 1}}}} - 1}}{{w - 1}},r_c^* = \alpha $ (If $0 < {x^*} < 1$, the fixed point is in domain of definition)

\begin{equation}
\begin{aligned}
J\left( {{x^*},\alpha } \right) = \left[ {\begin{array}{*{20}{c}}
A&B\\
0&D
\end{array}} \right]
\end{aligned}
  \end{equation}
where

\begin{eqnarray}
&& A = \frac{{{x^*}{\mkern 1mu} \left( {1 - {x^*}} \right)\left( {\alpha {\mkern 1mu} w - {r_d}} \right){\mkern 1mu} \left( {N - 1} \right){\mkern 1mu} {\mkern 1mu} {{\left( {w{\mkern 1mu} {x^*} - {x^*} + 1} \right)}^{N - 2}}}}{N}\notag\\
&& B = \frac{{{x^*}\left( {1 - {x^*}} \right){\mkern 1mu} \left( {w{\mkern 1mu} {{\left( {w{\mkern 1mu} {x^*} - {x^*} + 1} \right)}^{N - 1}} - 1} \right){\mkern 1mu} }}{{N{\mkern 1mu} \left( {w - 1} \right)}}\notag\\
&& D = \epsilon {\mkern 1mu} \left( {\beta  - \alpha } \right){\mkern 1mu} \left( { - {x^*}{\mkern 1mu} \left( {\frac{{\alpha \left( {w{\mkern 1mu} {{\left( {w{\mkern 1mu} {x^*} - {x^*} + 1} \right)}^{N - 1}} - 1} \right)}}{{N{\mkern 1mu} \left( {w - 1} \right)}} - 1} \right)} \right.\notag\\
&& \left. {+ \frac{{{\mkern 1mu} {r_d}\theta {\mkern 1mu} \left( {1 - {x^*}} \right){\mkern 1mu} \left( {{{\left( {w{\mkern 1mu} {x^*} - {x^*} + 1} \right)}^{N - 1}} - 1} \right)}}{{N{\mkern 1mu} \left( {w - 1} \right)}}} \right)
\end{eqnarray}

%  \begin{equation}
%\begin{aligned}

%\end{aligned}
%  \end{equation}

Eigenvalues are ${\lambda _1} = A$ and ${\lambda _2} = D$. Here we consider the  two following situations: (1) $w > 1$ and (2) $w < 1$.

Case $w > 1$: Due to ${r_d} < \alpha <w\alpha $, we have $\alpha w - {r_d}>0$, which indicates ${\lambda _1} > 0$. Therefore, this fixed point is unstable.

Case $w < 1$: Assume ${\lambda _1} < 0$. This assumption leads to $\alpha w - {r_d} < 0$. Here we set $N = 4$. Considering $0 < {x^*} = \frac{{{{\left( {\frac{{\alpha  - {r_d} + N{\mkern 1mu} \left( {w - 1} \right)}}{{\alpha w - {r_d}}}} \right)}^{\frac{{\rm{1}}}{{N - 1}}}} - 1}}{{w - 1}} < 1$, we can derive ${w^3} < \frac{{\alpha  - {r_d} + 4{\mkern 1mu} \left( {w - 1} \right)}}{{\alpha w - {r_d}}} < 1$. By calculating the right half of the inequality, we have $\alpha  - {r_d} + 4\left( {w - 1} \right) > \alpha w - {r_d}$, i.e. $w > 1$. Contradictory appears, so the assumption does not hold, which means that this fixed point is unstable.

(6) ${x^*} = \frac{{{{\left( {\frac{{\beta  - {r_d} + N{\mkern 1mu} \left( {w - 1} \right)}}{{\beta w - {r_d}}}} \right)}^{\frac{{\rm{1}}}{{N - 1}}}} - 1}}{{w - 1}},r_c^* = \beta $ (If $0 < {x^*} < 1$, the fixed point is in domain of definition)

Jacobian is similar to (5), one of whose eigenvalues is  $\lambda_1=A = \frac{{{x^*}{\mkern 1mu} \left( {1 - {x^*}} \right)\left( {\beta w - {r_d}} \right){\mkern 1mu} \left( {N - 1} \right){\mkern 1mu} {\mkern 1mu} {{\left( {w{\mkern 1mu} {x^*} - {x^*} + 1} \right)}^{N - 2}}}}{N} > 0$. Thus, this fixed point is unstable.

(7) Interior fixed point $\left( {{x^*} = \frac{\theta }{{\theta  + 1}},r_c^* = \frac{{N\left( {w - 1} \right) + {r_d}\left( {{{\left( {w{x^*} - {x^*} + 1} \right)}^{N - 1}} - 1} \right)}}{{w{{\left( {w{x^*} - {x^*} + 1} \right)}^{N - 1}} - 1}}} \right)$ (If $\alpha  \le r_c^* \le \beta $, the fixed point is in domain of definition)

  \begin{equation}
\begin{aligned}
J\left( {{x^*},r_c^*} \right) = \left[ {\begin{array}{*{20}{c}}
A&B\\
C&D
\end{array}} \right]
\end{aligned}
  \end{equation}
where
\begin{eqnarray}
&& A = \frac{{{\mkern 1mu} {x^*}\left( {1 - {x^*}} \right){\mkern 1mu} \left( {r_c^*w - {r_d}} \right){\mkern 1mu} \left( {N - 1} \right){\mkern 1mu} {\mkern 1mu} {{\left( {w{\mkern 1mu} {x^*} - {x^*} + 1} \right)}^{N - 2}}}}{N}\notag\\
&& B = \frac{{{x^{\rm{*}}}\left( {1 - {x^{\rm{*}}}} \right)\left( {w{\mkern 1mu} {{\left( {w{\mkern 1mu} {x^{\rm{*}}} - {x^{\rm{*}}} + 1} \right)}^{N - 1}} - 1} \right){\mkern 1mu} }}{{N{\mkern 1mu} \left( {w - 1} \right)}}\notag\\
&& C =  - {f_c}\left( {{x^*},r_c^*} \right) - \theta {f_d}\left( {{x^*},r_c^*} \right) - xf_c^{'}\left( {{x^*},r_c^*} \right) + \theta \left( {1 - x} \right)f_d^{'}\left( {{x^*},r_c^*} \right)\notag\\
&& =  - \left( {1 + \theta } \right){f_d}\left( {{x^*},r_c^*} \right) - xf_c^{'}\left( {{x^*},r_c^*} \right) + \theta \left( {1 - x} \right)f_d^{'}\left( {{x^*},r_c^*} \right)\notag\\
&& = {\mkern 1mu} \epsilon \left( {r_c^* - \alpha } \right)\left( {\beta  - r_c^*} \right)\left({\frac{{\left( {1 + \theta } \right){r_d}\left( {1 - {{\left( {w{x^*} - {x^*} + 1} \right)}^{N - 1}}} \right)}}{{N\left( {w - 1} \right)}}}\right.\notag\\
&& \left.{+ \left[ {{r_d}\theta \left( {1 - {x^*}} \right) - r_c^*w{x^*}} \right]\frac{{\left( {N - 1} \right){{\left( {w{x^*} - {x^*} + 1} \right)}^{N - 2}}}}{N}}\right)\notag\\
&& D = {\mkern 1mu} \epsilon {\mkern 1mu} \left( {r_c^* - \alpha } \right){\mkern 1mu} \left( {\beta  - r_c^*} \right)\frac{{{x^*}{\mkern 1mu} \left( {1 - w{\mkern 1mu} {{\left( {w{\mkern 1mu} {x^*} - {x^*} + 1} \right)}^{N - 1}}} \right){\mkern 1mu} }}{{N{\mkern 1mu} \left( {w - 1} \right)}}
\end{eqnarray}

This fixed point is stable if and only if the following two conditions holds:
  \begin{equation}
\begin{aligned}
\left\{ \begin{array}{l}
AD - BC > 0\\
A + D < 0
\end{array} \right.
\end{aligned}
  \end{equation}
The first equality
$AD - BC = \frac{{\epsilon \left( {r_c^* - \alpha } \right)\left( {\beta  - r_c^*} \right){x^*}\left( {1 - {x^*}} \right)}}{{{N^2}}}\frac{{w{{\left( {w{x^*} - {x^*} + 1} \right)}^{N - 1}} - 1}}{{w - 1}}\frac{{{r_d}\left( {1 + \theta } \right)\left( {{{\left( {w{x^{\rm{*}}} - {x^{\rm{*}}} + 1} \right)}^{N - 1}} - 1} \right)}}{{w - 1}} > 0$ always hold. Thus this point is stable if and only if $A + D > 0$, which is equivalent to the following condition:

  \begin{equation}
\begin{aligned}
\epsilon  > {\epsilon ^*}
\end{aligned}
  \end{equation}
where
  \begin{equation}
\begin{aligned}
{\epsilon ^*} &= \frac{{\left( {1 - {x^*}} \right)\left( {r_c^*w - {r_d}} \right)\left( {N - 1} \right){{\left( {w{x^*} - {x^*} + 1} \right)}^{N - 2}}\left( {w - 1} \right)}}{{\left( {r_c^* - \alpha } \right)\left( {\beta  - r_c^*} \right)\left[ {w{{\left( {w{x^*} - {x^*} + 1} \right)}^{N - 1}} - 1} \right]}}\\
&= \frac{{\left( {r_c^*w - {r_d}} \right)\left( {N - 1} \right){{\left( {w\theta  + 1} \right)}^{N - 2}}\left( {w - 1} \right)}}{{\left( {r_c^* - \alpha } \right)\left( {\beta  - r_c^*} \right)\left[ {w{{\left( {w\theta  + 1} \right)}^{N - 1}} - {{\left( {1 + \theta } \right)}^{N - 1}}} \right]}}
\end{aligned}
  \end{equation}

In conclusion, on the premise of $w \ne 1$, the stable fixed points and the conditions of stability are as follows:

(1) The fixed points on the vertical-axes $\left( x^{*} = 0 \right)$
are always stable.

(2) The fixed point $\left( {x^ {*} } = 1,r_c^ {*}  = \alpha \right)$ is stable when ${\mkern 1mu} \frac{{\alpha {\mkern 1mu} \left( {{w^N} - 1} \right)}}{{N{\mkern 1mu} \left( {w - 1} \right)}} > 1$ and ${r_d} < \frac{{\alpha \left( {{w^N} - 1} \right) - N\left( {w - 1} \right)}}{{\left( {{w^{N - 1}} - 1} \right)}}$.

(3) The interior fixed point $\left( {{x^*} = \frac{\theta }{{\theta  + 1}},r_c^* = \frac{{N\left( {w - 1} \right) + {r_d}\left( {{{\left( {w{x^*} - {x^*} + 1} \right)}^{N - 1}} - 1} \right)}}{{w{{\left( {w{x^*} - {x^*} + 1} \right)}^{N - 1}} - 1}}} \right)$ is stable when $\epsilon > \frac{{\left( {1 - {x^{*}}} \right)\left( {r_c^*w - {r_d}} \right)\left( {N - 1} \right){{\left( {w{x^{*}} - {x^{*}} + 1} \right)}^{N - 2}}\left( {w - 1} \right)}}{{\left( {r_c^{*} - \alpha } \right)\left( {\beta  - r_c^{*}} \right)\left[ {w{{\left( {w{x^{*}} - {x^{*}} + 1} \right)}^{N - 1}} - 1} \right]}}$.

\section*{Appendix B. The influence of multiplication factors of defectors $r_ {d}$ and initial phase $\delta$.}

\begin{figure}[!ht]
\includegraphics[width=0.95\linewidth]{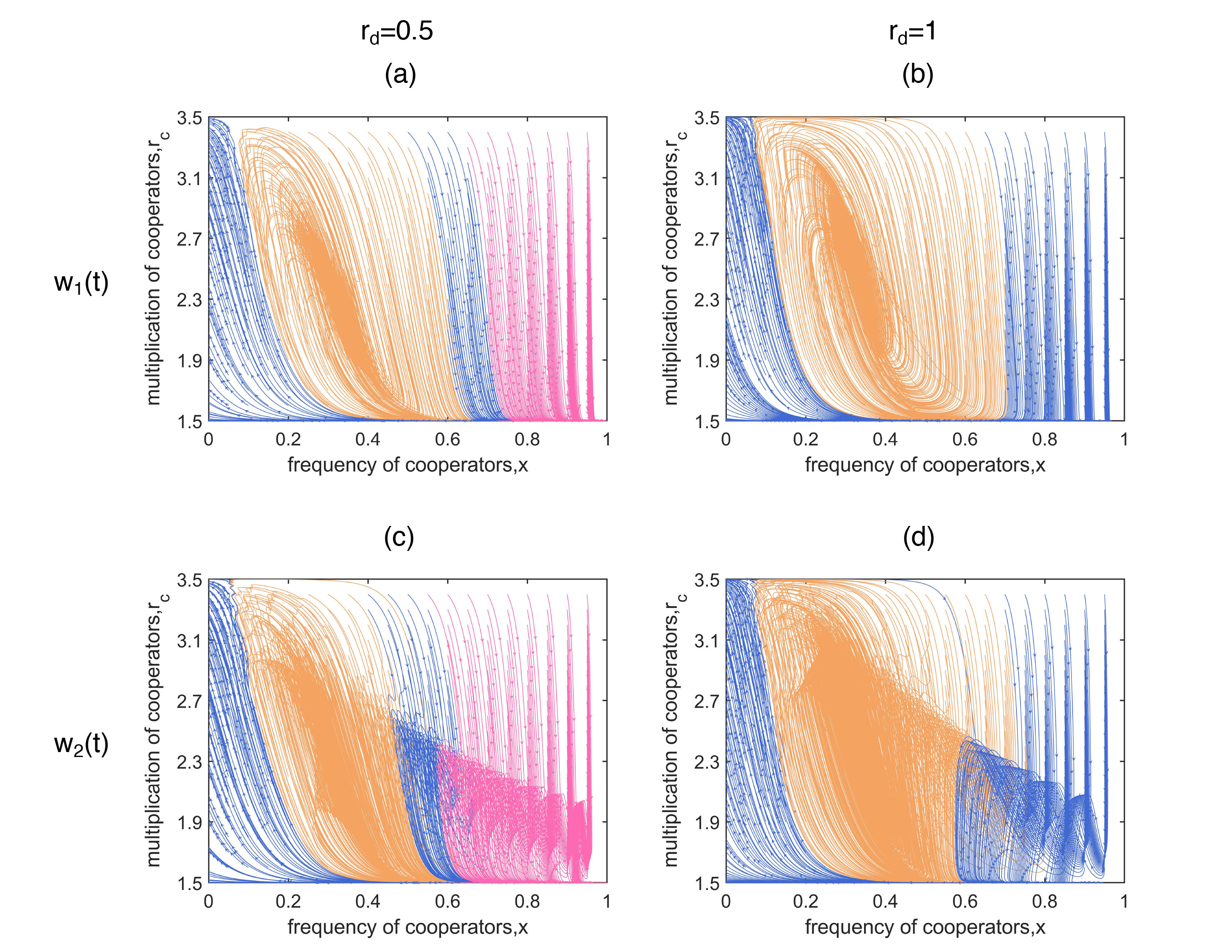}
\caption{{\bf Game-environment evolution under different multiplication factors of defector $\bm{r_d}$.}
$x-r_c$ phase plane shows that trajectories eventually evolve to $\left(x^{*}=0\right)$, $\left(x^{*}=1, r_{c}^{*}=\alpha\right)$ or circulate along an interior closed orbit, which are distinguished by blue, pink and orange, respectively. In all panels, $N=4$, $\alpha=1.5$, $\beta=3.5$, $\theta=0.5$, $a=1$, $\epsilon=6$, $\delta=0$.}
\label{fig:B1}
\end{figure}

In this part, we mainly aim to explore how multiplication factors of defectors $r_ {d}$ and initial phase $\delta$ affect the emergence of the interior closed orbit.

We first study the influence of $r_d$. Figure \ref{fig:B1} presents the phase plane of game-environment evolution showing trajectories of different initial points. Specifically, Figure \ref{fig:B1}(a)(b) describes situations in discretely varying global environment under different $r_d$, while Figure \ref{fig:B1}(c)(d) corresponds to continuously changing global environment. In each figure, there are always three kinds of trajectories represented by different colors, where orange lines represent trajectories evolving along an interior closed orbit. It indicates that the emergence of the interior closed orbit is robust on different $r_d$, whatever the environment is discretely varying or continuously changing. In addition, by comparing Figure \ref{fig:B1}(a) and (b) or Figure \ref{fig:B1}(c) and (b), we find that the value of $r_d$ indeed affect the range of orange region, i.e., the range of trajectories eventually evolving along the interior closed orbit.

\begin{figure}[!ht]
\includegraphics[width=0.95\linewidth]{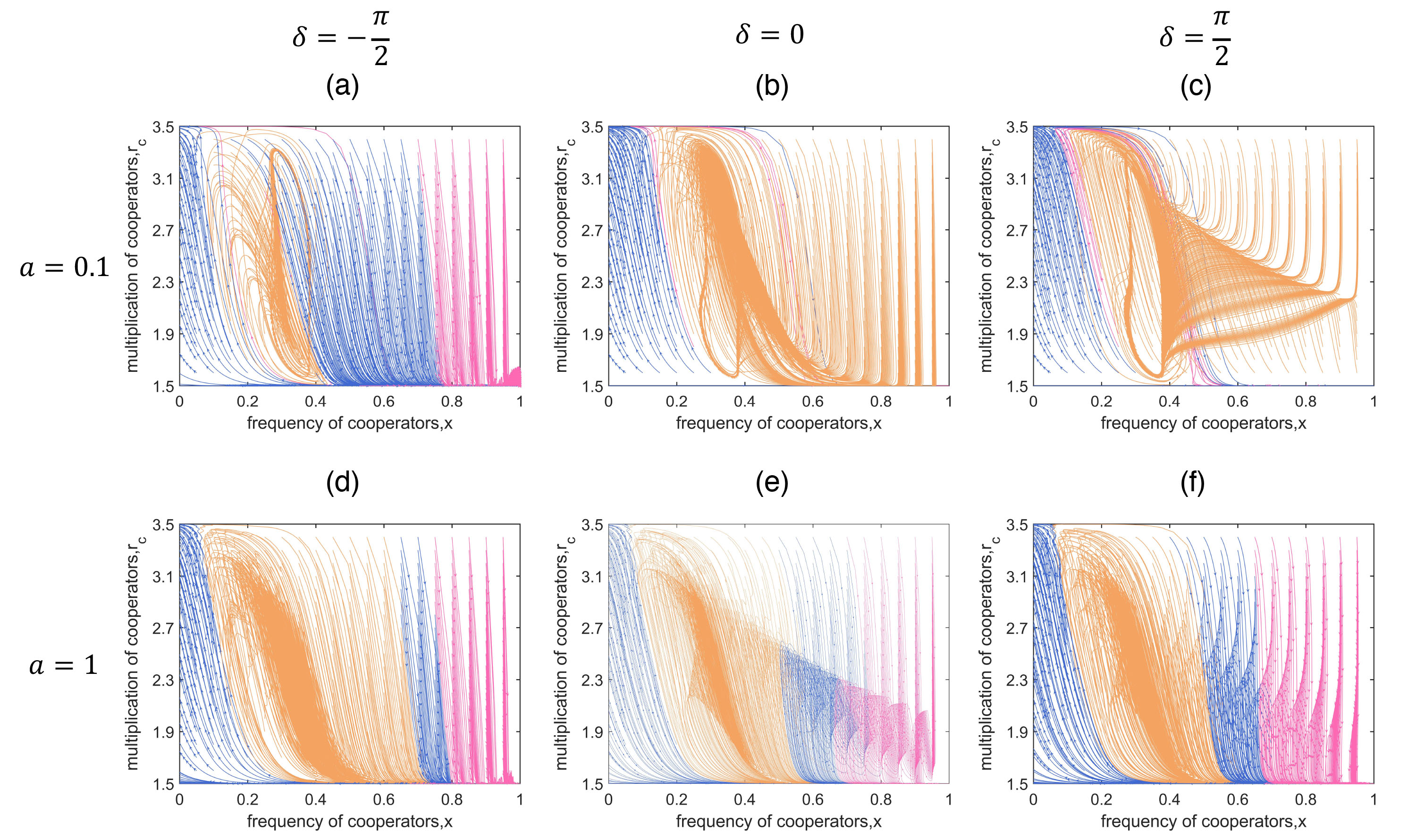}
\caption{{\bf Game-environment evolution under different initial phase $\bm{\delta}$ and self-changing speed $\bm{a}$ of continuously changing global environment.}
In $x-r_c$ plane, there are always three kinds of trajectories: the blue, pink and orange trajectories will eventually evolve to $\left(x^{*}=0\right)$,  $\left(x^{*}=1, r_{c}^{*}=\alpha\right)$ and circulate along a closed orbit, respectively. In all panels, $N=4$, $\alpha=1.5$, $\beta=3.5$, $\theta=0.5$, $r_d=0.6$,  $\epsilon=7$.}
\label{B2}
\end{figure}

Then, we explore the influence of $\delta$, the initial phase of the function $w_2(t)$ representing the continuously changing environment. Figure \ref{B2} presents $x-r_c$ phase plane under different $\delta$ and changing speed of environment $a$. Results show that all subfigures have a large range of orange trajectories eventually evolving along an interior closed orbit, although the shapes of orange range are distinguished in different figures. This indicates that the interior closed orbit always emerges for a large range of $\delta$ and that the value of $\delta$ can alter the attraction range of the  interior closed orbit.

\section*{Appendix C. Discretely changing global environment: four-segment function}

In this part, we consider a four-segment function $w_3(t)$ to describe the discretely changing global environments, which aims to check the explanations about why there emerges interior closed orbit. Specifically, the formula of $w_3(t)$ is 

\begin{equation}
\begin{aligned}
{w_{\rm{3}}}\left( t \right) = \left\{ {\begin{array}{*{20}{c}}
{1.4}&{\left[ t/T \right] = 4n}\\
{1}&{\left[ t/T  \right] = 4n+1}\\
{0.6}&{\left[ t/T  \right] = 4n+2}\\
{1}&{\left[ t/T  \right] = 4n+3}\\
\end{array}} \right.,n = 0,1,2, \cdots  \cdots
\end{aligned}
\end{equation}

\begin{figure}[!ht]
\includegraphics[width=0.95\linewidth]{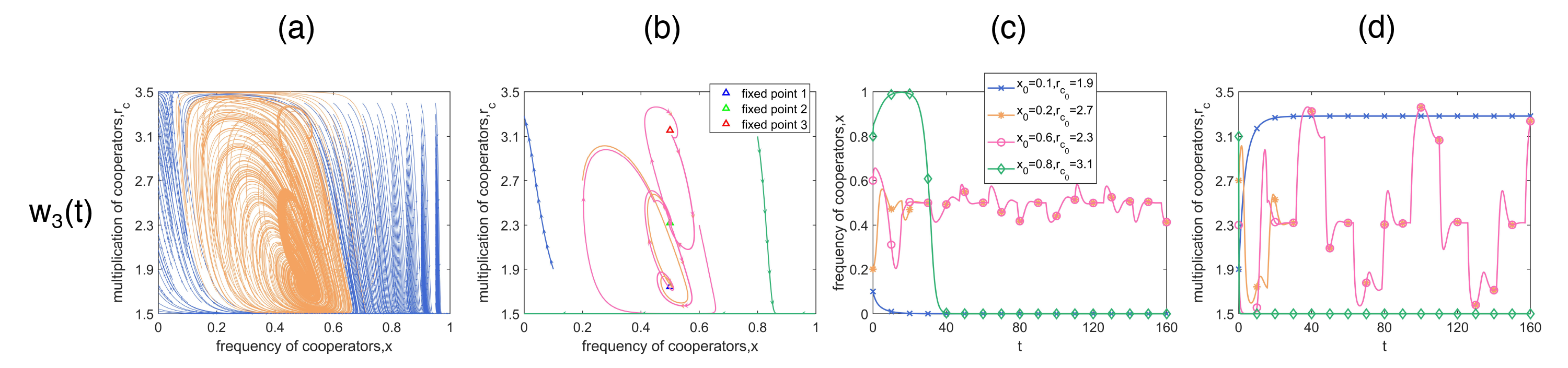}
\caption{{\bf Emergence of cyclic evolutions of group cooperation and local environment under periodically changing global environments given by $\bm{w_{3}\left(t\right)}$.}
 (a) Local game-environment evolutions. (b) Four typical dynamic trajectories with different initial conditions. (c-d) Time evolutions of $x$ and $r_c$, corresponding to the colored trajectories in (b). Parameters: $N=4$,$\alpha=1.5$, $\beta=3.5$, $\theta=1$, $r_{d}=1.2$, $a=0.1$, $\epsilon=4$.}
\label{S3}
\end{figure}

Figure \ref{S3} presents game-environment evolution under discretely changing environment given by $w_{3}\left(t\right)$. Results also show the emergence of cyclic evolutions (represented by orange trajectories in Figure \ref{S3}(a)). We present some representative trajectories through selecting different initial points in Figure \ref{S3}(b). Noteworthy, the orange and pink trajectories display the status of cyclic evolution of group cooperation and local environment. Take the orange trajectory as an example. The trajectory approaches fixed point 1 (labelled in figure \ref{S3}(b)) during the first $T$ time, then seized by fixed point 2 when the global environment changes to the second stage. When the global environment evolves to the third stage, unexpectedly, the dynamical trajectory is not actually captured but is still attracted by fixed point 3, which subsequently be seized by fixed point 2 again. Finally, as fixed point 2 is within the attract domain of fixed point 1, the trajectory forms a complete loop, leading to the cyclic evolution along an interior closed orbit consisting of four path segments. The time-evolution pictures (see Figure \ref{S3}(c)(d)) clearly present the periodic behaviors of $x$ and $r_c$. This complicated evolutionary process, which shows the important characteristics emerged both in discrete environment $w_1(t)$ where fixed point 1 and fixed point 2 are in the mutual attraction domains and in continuous environment $w_2(t)$ where the dynamical path are not captured yet are strongly attracted by the fixed point, supports our limit thought which explains the formation of the closed orbit under continuously changing environment to some extent. 

\end{document}